\documentclass[12pt,prd,tightenlines,nofootinbib,showpacs,showkeys]{revtex4}
\newcommand{\be}{\begin{equation}}
\newcommand{\ee}{\end{equation}}
\newcommand{\bdis}{\begin{displaymath}}
\newcommand{\edis}{\end{displaymath}}
\newcommand{\bga}{\begin{equation}\begin{gathered}}
\newcommand{\ega}{\end{gathered}\end{equation}}
\usepackage{bm}
\usepackage{graphics}
\usepackage{rotating}
\usepackage{epsfig}
\usepackage{amsmath}
\usepackage{amsfonts}

\begin{document}
\title{\begin{flushright}{\rm\normalsize SSU-HEP-14/05\\[5mm]}\end{flushright}
Pair double heavy diquark production in high energy proton-proton collisions}
\author{\firstname{A.P.} \surname{Martynenko}}
\affiliation{Samara State University, Pavlov Street 1, 443011, Samara, Russia}
\affiliation{Samara State Aerospace University named after S.P. Korolyov, Moskovskoye Shosse 34, 443086
Samara, Russia}
\author{\firstname{A.M.} \surname{Trunin}}
\affiliation{Samara State Aerospace University named after S.P. Korolyov, Moskovskoye Shosse 34, 443086
Samara, Russia}
\affiliation{Bogoliubov Laboratory of Theoretical Physics,
Joint Institute for Nuclear Research, Joliot-Curie Street 6, 141980 Dubna, Russia}

\begin{abstract}
On the basis of perturbative QCD and relativistic quark model we
calculate relativistic and bound state corrections in the production
processes of a pair of double heavy diquarks.
Relativistic factors in the production amplitude connected with the relative motion
of heavy quarks and the transformation law of the bound state wave function to the reference frame of the
moving $S$-wave diquark bound states are taken into account. For the gluon and quark propagators entering
the amplitudes we use a truncated expansion in relative quark momenta up to the second order.
Relativistic corrections to the quark-quark bound state wave functions in the rest
frame are considered by means of the Breit-like potential. It turns out that the examined effects
significantly decrease nonrelativistic cross sections.
\end{abstract}

\pacs{13.85.Ni, 12.38.Bx, 12.39.Ki}

\keywords{Hadron production in proton-proton interaction, Relativistic quark model}

\maketitle

\section{Introduction}

Double heavy meson and baryon production at high energies represents an important problem
of quantum chromodynamics (QCD). On the one hand the methods of nonrelativistic QCD (NRQCD) can be used in this case for a construction
of production amplitudes in the leading order over strong coupling constant $\alpha_s$ or for a calculation
of next-to-leading order corrections \cite{NRQCD}. On the other hand, the presence of heavy quarks gives a possibility to explore
the formation of quark bound states in these reactions on the basis of quark model. During last ten years in the problem
of double heavy hadron production there arises the field of research connected with pair double heavy quarkonium production.
The progress was initiated by experiments of Belle and BaBar collaborations which measured the cross sections of pair
charmonium production in $e^+e^-$ annihilation \cite{Belle,BaBar,brambilla,pahlova}. The importance of such reactions for a development
of theoretical methods of their investigation was demonstrated in~\cite{bodwin,chao,likhoded,ebert}. Essential improvement
in theoretical description of the processes of pair quarkonium production was obtained with the assumption of systematic
account of relativistic and radiative corrections to nonrelativistic results. It was revealed that corrections of relative
motion of heavy quarks and bound state corrections essentially change nonrelativistic calculations.

Another reaction of
pair meson production was investigated recently in $pp$ interaction \cite{LHCb}. Beginning with the start of the LHC activity,
new experimental data on double heavy quarkonium production regenerated the interest to the study of quarkonium production
mechanisms in hadronic collisions. The production of $J/\psi J/\psi$ pair in
proton-proton collisions at a center-of-mass energy $\sqrt{s}=7$~TeV has been observed with the LHCb detector.
The data used for the analysis was obtained with an integrated luminosity of
37.5~$pb^{-1}$ of $pp$ collisions at a center-of-mass
energy of $\sqrt{s}$ = 7 TeV collected by the LHCb experiment in 2010.
At collider energies, double charmonium production occurs for the most part through
$gluon-gluon$ channel. Theoretical description of pair charmonium production was carried out in this case in nonrelativistic
approximation in~\cite{likhoded1988,likhoded1,qiao,ko} and with the account of relativistic corrections in~\cite{mt2012,xu}.
Note that additional uncertainty occurs in $pp$ interaction due to double parton scattering mechanism \cite{baranov,baranov1}.
Along with pair quarkonium production there is the interest to pair double heavy diquark production because such
process can represent a first stage of double baryon production \cite{baryon}.
The pair production of double heavy diquarks in $e^+e^-$ and
$p\bar p$ interaction was performed in nonrelativistic QCD in~\cite{braguta}. An account of relativistic and bound state corrections
to the cross sections in the case of $e^+e^-$ annihilation was carried out in~\cite{mt2014}. It was shown in~\cite{mt2014} that
a reliable estimate of observed cross sections can be obtained only with systematic account of relativistic and
bound state corrections. It worth mentioning that the pair diquark $(cc)$
and $(\bar c\bar c)$ production at the LHC energies with subsequent formation of a tetra-quark was studied in~\cite{likhoded1}.
In this work we continue an investigation of relativistic effects in pair double heavy diquark production
in proton-proton interaction at energies of the LHC. We calculate cross section $\sigma(pp\to {\cal D}\bar {\cal D}+X)$
of pair diquark production in nonrelativistic approximation and show how these results will be changed after the account of relativistic
corrections.

\section{General formalism}

All the models describing quarkonium production in hadronic collisions use the common basis:
the factorisation between the hard collison subprocess and the parton-parton collision
luminosity, calculated as a convolution of the parton distribution functions (PDFs).
In collinear parton model the cross section of pair double heavy diquark production in proton-proton collisions has the form
of the convolution of partonic cross section \hbox{$d\sigma[gg\to D_{bc} + \bar D_{\bar b \bar c}]$} with the parton
distribution functions of initial protons~\cite{likhoded1988,braaten,kramer}:
\be
\label{eq:cs-plus-x}
d\sigma[p+p\to D_{bc} + \bar D_{\bar b \bar c}+X]=\int \! dx_1 dx_2 \, f_{g/p}(x_1,\mu) f_{g/p}(x_2,\mu)
\, d\sigma[gg\to D_{bc} + \bar D_{\bar b \bar c}],
\ee
where $f_{g/p}(x,\mu)$ is the parton (gluon) distribution function (PDF) in the proton, $x_{1,2}$ are the
parton momentum (longitudinal momentum) fraction from the proton, $\mu$ is the factorization scale.
Neglecting the proton mass and taking the
c.m. reference frame of initial protons with the beam along the $z$-axis
we can present the gluon on mass-shell momenta as $k_{1,2}=x_{1,2}\frac{\sqrt{S}}{2}(1,0,0,\pm 1)$.
$\sqrt{S}$ is the center-of-mass energy in proton-proton collision. The range of accessible $x_{1,2}$
depends on the rapidity interval covered by experiments. At the CM energies of the LHC the gluon-gluon
contribution to the production cross section is dominant, so that we consider only $gg$ initial states in
this study. Quark-antiquark annihilation amounts to about 10\% \cite{likhoded1}

According to the quasipotential approach the double heavy diquark production amplitude for the gluonic
subprocess $gg\to D_{bc} + \bar D_{\bar b \bar c}$ can be expressed as a convolution of a perturbative production
amplitude of $(bc)$ and $(\bar b \bar c)$ quark and anti-quark pairs $\mathcal T(p_1,p_2;q_1,q_2)$ and the
quasipotential wave functions of final diquarks $\Psi_{\mathcal D}$~\cite{ebert,mt2012}:
\be
\label{eq:m-gen}
{\mathcal M}[gg\to D_{bc} + \bar D_{\bar b \bar c}](k_1,k_2,P,Q)=\int \! \frac{d\mathbf p}{(2\pi)^3}
\int \! \frac{d\mathbf q}{(2\pi)^3} \, \bar\Psi_{D_{bc}}(p,P) \bar\Psi_{\bar D_{\bar b \bar c}}(q,Q) \otimes \mathcal T(p_1,p_2;q_1,q_2),
\ee
where $p_{1,2}$ are four-momenta of $c$ and $b$ quarks, and $q_{1,2}$ are an appropriate four-momenta of $\bar c$ and $\bar b$ anti-quarks.
They are defined in terms of total momenta $P(Q)$ and relative momenta $p(q)$ as follows:
\bga
p_{1,2}=\eta_{1,2} P \pm p,\quad (pP)=0; \qquad q_{1,2}=\eta_{1,2} Q \pm q,\quad (qQ)=0, \\
\eta_{1,2} = \frac{M^2 \pm m_c^2 \mp m_b^2}{2M^2},
\ega
where $M = M_{D_{bc}} = M_{\bar D_{\bar b \bar c}} $ is the double heavy diquark mass, $p=L_P(0,\mathbf p)$ and $q=L_Q(0,\mathbf q)$
are the relative four-momenta obtained by the Lorentz transformation of four-vectors
$(0,\mathbf p)$ and $(0,\mathbf q)$ to the reference frames moving with the four-momenta $P$~and~$Q$ of final diquarks $D_{bc}$
and $\bar D_{\bar b \bar c}$. In Eq.~\eqref{eq:m-gen} we integrate over the relative three-momenta of quarks and antiquarks
in the final state. The wave functions $\bar\Psi_{D_{bc}}(p,P)$ and $\bar\Psi_{\bar D_{\bar b \bar c}}(q,Q)$ determine the
probability for free heavy quark $Q_1Q_2$ and anti-quark $\bar Q_1\bar Q_2$ pairs with certain quantum numbers to transform into
diquark and anti-diquarks bound states (long distance matrix elements). A proof of factorization formulas~\eqref{eq:cs-plus-x}-\eqref{eq:m-gen}
deserves a special consideration. There are interactions between initial and final hadrons
connected with gluon exchanges that violate this factorization. In what follows we assume that the emission of soft and collinear gluons
can be absorbed into parton distribution functions and long distance matrix elements, so that the factorization
equations~\eqref{eq:cs-plus-x}-\eqref{eq:m-gen} occur. The status of a proof of factorization in quarkonium
production is presented in detail in \cite{gtb}. A proof of factorization is essential because non-factorizing gluon contributions,
for which $\alpha_s$ is not small, could change numerical results.
It should be mentioned that the effect induced by the radiation in the initial state was investigated in \cite{likhoded1}
by means of Pythia Monte Carlo generator. It does not effect the value of total cross section.

The parton-level differential cross section for $g+g\to D_{bc}+\bar D_{\bar b\bar c}$ is expressed further through
the Mandelstam variables $s$, $t$ and $u$:
\be
\label{eq:stu-def}
s=(k_1+k_2)^2=(P+Q)^2=x_1x_2S,
\ee
\bdis
t=(P-k_1)^2=(Q-k_2)^2=M^2-x_1\sqrt{S}(P_0-|{\bf P}|\cos\phi)=M^2-x_1x_2S+x_2\sqrt{S}(P_0+|{\bf P}|\cos\phi),
\edis
\be
u=(P-k_2)^2=(Q-k_1)^2=M^2-x_2\sqrt{S}(P_0+|{\bf P}|\cos\phi)=M^2-x_1x_2S+x_1\sqrt{S}(P_0-|{\bf P}|\cos\phi),
\ee
where $\phi$ is the angle between ${\bf P}$ and the $z$-axis. The Mandelstam variables $s$, $t$ and $u$ satisfy
to relation
\be
\label{eq:stu}
s+t+u=M^2_{D_{bc}}+M^2_{\bar D_{\bar b\bar c}}.
\ee
The transverse momentum $P_T$ of diquark $D_{bc}$ and its energy $P_0$ can be written as
\be
\label{eq:pt0-def}
P_T^2=|{\bf P}|^2\sin^2\!\phi=-t-\frac{(M^2-t)^2}{x_1x_2S},\quad P_0=\frac{x_1x_2\sqrt{S}}{x_1+x_2}+\frac{x_1-x_2}{x_1+x_2}|{\bf P}|\cos\phi.
\ee

\begin{figure}[t!]
\center\includegraphics[scale=0.75]{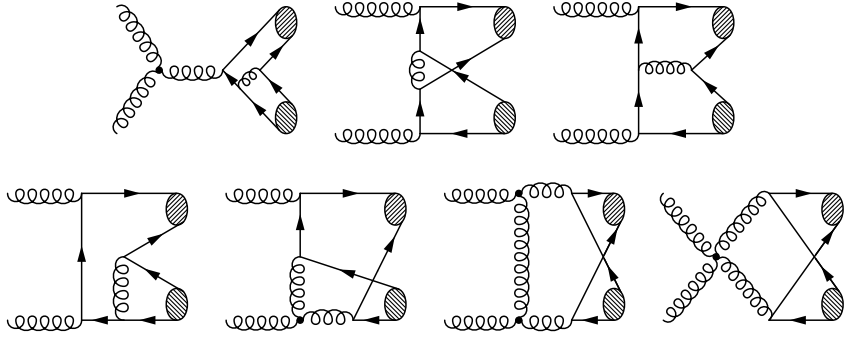}
\caption{The typical leading order diagrams
for $gg\to D_{bc} + \bar D_{\bar b \bar c}$ subprocess.
Other diagrams can be obtained by reversing the quark lines or interchanging the initial gluons.
}
\label{fig:d35}
\end{figure}

\begin{figure}[t!]
\center\includegraphics[scale=0.75]{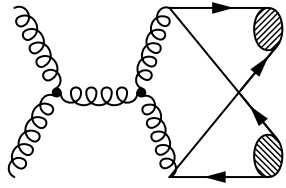}
\caption{The additional diagram
for $gg\to D_{bc} + \bar D_{\bar b \bar c}$ having the zero color factor.
}
\label{fig:d1}
\end{figure}

In the leading order in strong coupling constant $\alpha_s$, there are the 35 Feynman
diagrams contributing to gluon fusion subprocess \hbox{$gg\to D_{bc} + \bar D_{\bar b \bar c}$} of pair
double heavy diquark production,
which are presented in Fig.~\ref{fig:d35}. One additional diagram shown in Fig.~\ref{fig:d1} appears to have zero color factor
after summation with antisymmetric color functions $\epsilon^{ijk}/\sqrt2$ of final diquark states:
$f^{g_1g_2e}f^{abe}T^a_{c_1c_3}T^b_{c_2c_4}\epsilon^{c_1c_2A}\epsilon^{c_3c_4B}=0$, because we have here a convolution
of anti-symmetric $f^{abe}$ and symmetric over indices $a,b$ tensors.
$f^{abc}$ are the structure constants of SU(3) color group, $T^a$ is the SU(3) generator in the fundamental representation.
In view of large volume of calculations we have used the package FeynArts~\cite{feynarts}
for the system Mathematica in order to obtain analytical expressions for all the diagrams and, subsequently, Form~\cite{form}
to evaluate their traces. Then we obtain the following result for the leading order production amplitude~\eqref{eq:m-gen}:
 \begin{equation}
 \label{eq:mmm}
\mathcal M[gg\to D_{bc} + \bar D_{\bar b \bar c}](k_1,k_2,P,Q)=M\pi^2\alpha_s^2\int\!\frac{d\mathbf p}{(2\pi)^3}
\int\!\frac{d\mathbf q}{(2\pi)^3} \mathrm{Tr}\,\mathfrak M,
 \end{equation}
 \bdis
\mathfrak M= \bar\Psi_{P,p}^{bc}\gamma_\beta\bar\Psi_{Q,q}^{cb}\gamma_\omega\Gamma_1^{\beta\omega}+
\bar\Psi_{P,-p}^{cb}\gamma_\beta\,\Gamma_2^{\beta\omega\theta} \gamma_\omega\bar\Psi_{Q,-q}^{bc}\gamma_\theta+\bar\Psi_{P,p}^{bc}\gamma_\beta\,\Gamma_3^{\beta\omega\theta}
\gamma_\omega\bar\Psi_{Q,q}^{cb}\gamma_\theta
 \edis
 \bdis +
\bar\Psi_{P,p}^{bc}\,\hat\varepsilon_1\frac{m_c-\hat k_1+\hat p_1}{(k_1-p_1)^2-m_c^2}\gamma_\beta(\bar\Psi_{Q,q}^{cb}\gamma_\omega\Gamma_4^{\beta\omega}+\Gamma_5^{\beta\omega}
\bar\Psi_{Q,q}^{cb}\gamma_\omega)
 \edis
 \bdis
+
\bar\Psi_{P,-p}^{cb}\,\hat\varepsilon_1\frac{m_b-\hat k_1+\hat p_2}{(k_1-p_2)^2-m_b^2}
\gamma_\beta(\bar\Psi_{Q,-q}^{bc}\gamma_\omega\Gamma_6^{\beta\omega}+
\Gamma_7^{\beta\omega}\bar\Psi_{Q,-q}^{bc}\gamma_\omega)
 \edis
 \bdis
+\bar\Psi_{P,p}^{bc}\,\hat\varepsilon_2\frac{m_c-\hat k_2+\hat p_1}{(k_2-p_1)^2-m_c^2}
\gamma_\beta\Gamma_8^{\beta\omega}\bar\Psi_{Q,q}^{cb}\gamma_\omega
+\bar\Psi_{P,-p}^{cb}\,\hat\varepsilon_2\frac{m_b-\hat k_2+\hat p_2}{(k_2-p_2)^2-m_b^2}
\gamma_\beta\Gamma_9^{\beta\omega}\bar\Psi_{Q,-q}^{bc}\gamma_\omega
 \edis
\begin{equation}
\label{eq:m-main}
+\bar\Psi_{P,-p}^{cb}\gamma_\beta\frac{m_b+\hat k_1-\hat q_2}{(k_1-q_2)^2-m_b^2}\hat\varepsilon_1
\bar\Psi_{Q,-q}^{bc}\gamma_\omega\Gamma_{10}^{\beta\omega}
+\bar\Psi_{P,p}^{bc}\gamma_\beta\frac{m_c+\hat k_1-\hat q_1}{(k_1-q_1)^2-m_c^2}\hat\varepsilon_1
\bar\Psi_{Q,q}^{cb}\gamma_\omega\Gamma_{11}^{\beta\omega},
\end{equation}
where $\varepsilon_{1,2}$ are polarization vectors of initial gluons, the hat symbol means contraction of the four-vector
with the Dirac gamma-matrices. A number of vertex functions $\Gamma_i$ is introduced  to make the entry of the
amplitude~\eqref{eq:m-main} more compact. We explicitly extracted in~\eqref{eq:m-main} the normalization factors
$\sqrt{2M}$ of the quasipotential bound state wave functions.

The formation of diquark states from quark and anti-quark
pairs, which corresponds to first stage of double heavy baryon production, is determined in the quark model by the
quasipotential wave functions $\Psi_{D_{bc}}(p,P)$ and
$\Psi_{\bar D_{\bar b \bar c}}(q,Q)$. These wave functions are calculated initially
in the meson rest frame and then transformed to the reference frames moving with the four-momenta
$P$ and $Q$. The law of such transformation was derived in the Bethe-Salpeter approach in ~\cite{brodsky2}
and in the quasipotential method in~\cite{faustov}. We use the last one and obtain the following
expressions for the relativistic wave functions~\cite{mt2014}:
\begin{equation}
\begin{gathered}
\label{eq:relwf}
\bar\Psi_{D_{bc}}(p,P)=\frac{\bar\Psi_{D_{bc}}^{0}(\mathbf p)}{\sqrt{ \frac{\epsilon_c(p)}{m_c}
\frac{\epsilon_c(p)+m_c}{2m_c} \frac{\epsilon_b(p)}{m_b} \frac{\epsilon_b(p)+m_b}{2m_b} }}
\left[\frac{\hat v_1-1}{2}+\hat v_1\frac{\mathbf p^2}{2m_b(\epsilon_b(p)+m_b)}-\frac{\hat p}{2m_b}
\right] \\ \times
\Sigma^P(1+\hat v_1)
\left[\frac{\hat v_1+1}{2}+\hat v_1\frac{\mathbf p^2}{2m_c(\epsilon_c(p)+m_c)}+\frac{\hat p}{2m_c}
\right],
\\
\bar\Psi_{D_{\bar c\bar b}}(q,Q)=\frac{\bar\Psi_{\bar D_{\bar b \bar c}}^{0}(\mathbf q)}
{\sqrt{ \frac{\epsilon_c(q)}{m_c} \frac{\epsilon_c(q)+m_c}{2m_c} \frac{\epsilon_b(q)}{m_b} \frac{\epsilon_b(q)+m_b}{2m_b} }}
\left[\frac{\hat v_2-1}{2}+\hat v_2\frac{\mathbf q^2}{2m_c(\epsilon_c(q)+m_c)}+\frac{\hat q}{2m_c}
\right] \\ \times \Sigma^Q(1+\hat v_2)
\left[\frac{\hat v_2+1}{2}+\hat v_2\frac{\mathbf q^2}{2m_b(\epsilon_b(q)+m_b)}-\frac{\hat q}{2m_b}
\right],
\end{gathered}
\end{equation}
where $m_{c,b}$ are the quark masses, $\epsilon_{c,b}(p)=\sqrt{p^2+m_{c,b}^2}$, $v_1 = P/M$, $v_2 = Q/M$,
and $\Sigma^{P,Q}$ are equal to $\gamma_5$ and $\hat\varepsilon_{P,Q}$ for scalar and axial-vector diquarks
respectively. The polarization vectors $\varepsilon_{P,Q}$ of axial-vector diquarks satisfy the conditions:
$(\varepsilon_{P}\cdot P) = 0$ and $(\varepsilon_{Q}\cdot Q) = 0$. Quasipotential wave
functions~\eqref{eq:relwf} include projection operators on the states with definite spins:
$\bar u_i(0)\bar u_j(0)=[C\hat\varepsilon(\gamma_5)(1+\gamma_0)]_{ij}/2\sqrt2$ and
$v_i(0)v_j(0)=[(1-\gamma_0)\hat\varepsilon(\gamma_5)C]_{ij}/2\sqrt2$, where $C$ is the charge conjugation matrix.

Leading order vertex functions $\Gamma_i$ in~\eqref{eq:m-main} have the following form:
\bdis
\Gamma_1^{\beta\omega}=
{\mathcal K_1} D_{\mu}{}^{\beta}(p_1+q_1)D_{\nu}{}^{\omega}(p_2+q_2)\bigl(\varepsilon_1^\nu \varepsilon_2^\mu+\varepsilon_1^\mu \varepsilon_2^\nu-2g^{\mu\nu}(\varepsilon_1 \varepsilon_2)
 \edis
 \bdis
-D_{\lambda\kappa}(k_1-p_1-q_1)\mathfrak{E}_1^{\lambda\mu}(p_1+q_1)\mathfrak{E}_2^{\kappa\nu}(p_2+q_2) -
D_{\kappa\lambda}(k_1-p_2-q_2)\mathfrak{E}_1^{\kappa\nu}(p_2+q_2)\mathfrak{E}_2^{\lambda\mu}(p_1+q_1)\bigr),
 \edis
 \bdis
\Gamma_2^{\beta\omega\theta}={\mathcal K_{2}}\mathfrak{E}_2^{\mu}(-k_1) D_{\mu}{}^{\beta}(k_1+k_2)
D^{\theta\omega}(p_1+q_1)\frac{m_b-\hat p_1-\hat q_1-\hat q_2}{(p_1+q_1+q_2)^2-m_b^2}
 \edis
 \bdis +
{\mathcal K_{5}}\varepsilon_2^\omega \mathfrak{E}_1^{\mu\nu}(p_1+q_1) D_{\mu}{}^{\beta}(k_1-p_1-q_1)
D_{\nu}{}^{\theta}(p_1+q_1) \frac{m_b+k_2-q_2}{(k_2-q_2)^2-m_b^2}
 \edis
 \bdis +
D^{\theta\beta}(p_1+q_1)\frac{m_b+\hat p_1+\hat p_2+\hat q_1}{(p_1+p_2+q_1)^2-m_b^2} \Bigr(
{\mathcal K_{2}}\mathfrak{E}_1^{\mu}(-k_2) D_{\mu}{}^{\omega}(k_1+k_2)
+{\mathcal K_{9}}\varepsilon_1^\omega\hat\varepsilon_2\frac{m_b+\hat k_1-\hat q_2}{(k_1-q_2)^2-m_b^2}
 \edis
 \bdis
+{\mathcal K_{7}}\varepsilon_2^\omega\hat\varepsilon_1\frac{m_b+\hat k_2-\hat q_2}{(k_2-q_2)^2-m_b^2} \Bigr),
 \edis
 \bdis
\Gamma_4^{\beta\omega}={\mathcal K_3}D^{\beta\omega}(k_1-p_1-q_1)\frac{m_b+\hat k_2-\hat p_2}{(k_2-p_2)^2-m_b^2}\hat\varepsilon_2 -
{\mathcal K_4}\varepsilon_2^\omega D^{\beta\mu}(k_1-p_1-q_1)\frac{m_b-\hat k_2+\hat q_2}{(k_2-q_2)^2-m_b^2}\gamma_\mu
 \edis
 \bdis -
{\mathcal K_5}\mathfrak{E}_2^{\mu\nu}(p_2+q_2)D_{\mu}{}^{\beta}(k_1-p_1-q_1)D_{\nu}{}^{\omega}(p_2+q_2) ,
 \edis
 \bdis
\Gamma_5^{\beta\omega}={\mathcal K_6}D^{\beta\omega}(p_2+q_2)\frac{m_c+\hat k_2-\hat q_1}{(k_2-q_1)^2-m_c^2}\hat\varepsilon_2 +
{\mathcal K_7}\varepsilon_2^\beta D_{\mu}{}^{\omega}(p_2+q_2)\frac{m_c-\hat p_2-\hat q_1-\hat q_2}{(p_2+q_1+q_2)^2-m_c^2}\gamma_\mu,
 \edis
 \bdis
\Gamma_8^{\beta\omega}={\mathcal K_8}D^{\beta\omega}(p_2+q_2)\frac{m_c+\hat k_1-\hat q_1}{(k_1-q_1)^2-m_c^2}\hat\varepsilon_1
+ {\mathcal K_9}\varepsilon_1^\beta D_{\mu}{}^{\omega}(p_2+q_2)\frac{m_c-\hat p_2-\hat q_1-\hat q_2}{(p_2+q_1+q_2)^2-m_c^2}\gamma_\mu
 \edis
 \bdis
-{\mathcal K_{10}}\mathfrak{E}_1^{\mu\nu}(p_2+q_2)D_{\mu}{}^{\beta}(k_1-p_2-q_2)D_{\nu}{}^{\omega}(p_2+q_2),
 \edis
\bdis
\Gamma_{10}^{\beta\omega}={\mathcal K_{11}}D^{\beta\omega}(k_1-p_2-q_2)\frac{m_c+\hat k_2-\hat p_1}{(k_2-p_1)^2-m_c^2}\hat\varepsilon_2+{\mathcal K_{3}}\varepsilon_2^\omega D^{\beta\mu}(k_1-p_2-q_2)\frac{m_c-\hat k_2+\hat q_1}{(k_2-q_1)^2-m_c^2}\gamma_\mu
\edis
\begin{equation}
\label{eq:gammas}
+ {\mathcal K_{10}}\mathfrak{E}_2^{\mu\nu}(p_1+q_1)D_{\mu}{}^{\beta}(k_1-p_2-q_2)D_{\nu}{}^{\omega}(p_1+q_1),
\end{equation}
where we introduce the following tensors:
\begin{equation}
\begin{gathered}
\mathfrak{E}_{1,2}^{\mu\nu}(x)=g^{\mu\nu}(k_{1,2}-2x)\varepsilon_{1,2}+\varepsilon_{1,2}^\mu(2k_{1,2}^\nu-x^\nu)+\varepsilon_{1,2}^\nu(k_{1,2}^\mu+x^\mu), \quad \mathfrak{E}_{1,2}^{\mu}(x)=\varepsilon_{2,1}^\nu\mathfrak{E}_{1,2}^{\mu\nu}(x),
\end{gathered}
\end{equation}
and $D_{\mu\nu}(k)$ is the gluon propagator which is taken in the Feynman gauge.
Other vertex functions $\Gamma_i$ can be obtained by means of simultaneous replacement
$m_c \leftrightarrow m_b$, $p_1 \leftrightarrow p_2$, and $q_1 \leftrightarrow q_2$ in Eqs.~\eqref{eq:gammas}:
\begin{equation}
\begin{gathered}
\Gamma_3^{\beta\omega\theta}=\Gamma_2^{\beta\omega\theta}\biggl|{}_{\substack{m_b
\rightleftharpoons m_c \\ p_1 \rightleftharpoons p_2 \\ q_1 \rightleftharpoons q_2}}, \quad
\Gamma_6^{\beta\omega}=\Gamma_4^{\beta\omega}\biggl|{}_{\substack{m_b \rightleftharpoons m_c \\ p_1
\rightleftharpoons p_2 \\ q_1 \rightleftharpoons q_2}}, \quad
\Gamma_7^{\beta\omega}=\Gamma_5^{\beta\omega}\biggl|{}_{\substack{m_b \rightleftharpoons m_c \\ p_1
\rightleftharpoons p_2 \\ q_1 \rightleftharpoons q_2}}, \\
\Gamma_9^{\beta\omega}=\Gamma_8^{\beta\omega}\biggl|{}_{\substack{m_b \rightleftharpoons m_c \\ p_1
\rightleftharpoons p_2 \\ q_1 \rightleftharpoons q_2}}, \quad
\Gamma_{11}^{\beta\omega}=\Gamma_{10}^{\beta\omega}\biggl|{}_{\substack{m_b \rightleftharpoons m_c \\ p_1
\rightleftharpoons p_2 \\ q_1 \rightleftharpoons q_2}}. \quad
\end{gathered}
\end{equation}

Color factors of the Feynman amplitudes should be contracted over color indices with antisymmetric color
functions $\epsilon^{c_1c_2A}/\sqrt2$
and $\epsilon^{c_3c_4B}/\sqrt2$ ($c_i,A,B = 1,2,3$) of $D_{bc}$ and $\bar D_{\bar b \bar c}$ diquarks.
As a result we obtain the 11 different color factors $\mathcal K_i$ in~\eqref{eq:gammas}, which can be presented as follows:
\bdis
{\mathcal K_1}=-3{\mathcal C_0}-3{\mathcal C_1}+4{\mathcal C_3}, \quad {\mathcal K_2}=\frac43{\mathcal C_1}, \quad
{\mathcal K_3}=\frac{2i}3({\mathcal C_0}+2{\mathcal C_1}-4{\mathcal C_2}),
 \edis
 \bdis
{\mathcal K_4}=\frac{i}3({\mathcal C_0}-{\mathcal C_1}-{\mathcal C_2}), \quad
{\mathcal K_5}=\frac32{\mathcal C_0}+{\mathcal C_1}-2{\mathcal C_3}, \quad
{\mathcal K_6}=-\frac{i}{3}({\mathcal C_0}+3{\mathcal C_1}-5{\mathcal C_2}),
 \edis
 \bdis
{\mathcal K_7}=\frac{2i}{3}({\mathcal C_0}-2{\mathcal C_2}), \quad
{\mathcal K_8}=-\frac{i}3({\mathcal C_0}+2{\mathcal C_1}-5{\mathcal C_2}), \quad
{\mathcal K_9}=\frac{2i}3({\mathcal C_0}+2{\mathcal C_1}-2{\mathcal C_2}),  \quad
 \edis
 \bdis
{\mathcal K_{10}}=\frac32{\mathcal C_0}+2{\mathcal C_1}-2{\mathcal C_3},  \quad
{\mathcal K_{11}}=-\frac{i}3({\mathcal C_0}+2{\mathcal C_1}-{\mathcal C_2}),
 \edis
\be
\mathcal C_0=\delta^{g_1g_2}\delta_{AB}, \quad
\mathcal C_1=if^{g_1g_2a}(T^a)_{BA}, \quad
\mathcal C_2=(T^{g_1}T^{g_2})_{BA}, \quad
\mathcal C_3=f^{g_1ea}f^{g_2eb}(T^aT^b)_{BA},
\ee
where $g_{1,2} = 1, \ldots, 8$ are the color indices of initial gluons, $A$ and $B$ are the color indices
of final diquarks.

Let us present here, for example, the transformation of the first amplitude in Fig.~\ref{fig:d35}
from $\mathcal T_1(p_1,p_2;q_1,q_2)$ in \eqref{eq:m-gen} to $\mathcal M_1(k_1,k_2;P,Q)$ in \eqref{eq:mmm}
which takes the form in the Feynman gauge:
\bga
\mathcal T_1(p_1,p_2;q_1,q_2)=-
8i\,\pi^2\alpha_s^2\,
f^{g_1g_2b}(T^a)_{c_1c_3}(T^b)_{c_2c}(T^a)_{cc_4} e^{c_1c_2A}e^{c_3c_4B}
[\bar u(p_1)\gamma_\alpha v(q_1)]\\
\times
\Bigl[\bar u(p_2)\gamma_\beta\frac{m_b-\hat p_1-\hat q_1-\hat q_2}{(p_1+q_1+q_2)-m_b^2}\gamma_\omega v(q_2)\Bigr]
\varepsilon_1^\mu(k_1)\varepsilon_2^\nu(k_2) D^{\alpha\omega}(p_1+q_1)D^{\rho\beta}(k_1+k_2)
\\
\times
\bigl(g_{\mu\nu}(k_2-k_1)_\rho-g_{\nu\rho}(k_1+2k_2)_\mu+g_{\mu\rho}(2k_1+k_2)_\nu\bigr),
\ega

\begin{equation}
\begin{gathered}
\label{p1:eq:m-amp}
\mathcal M_1(k_1,k_2;P,Q)=-\frac43i\,\pi^2\alpha_s^2\sqrt{M_{D_{bc}}M_{\bar D_{\bar b\bar c}}}\, f^{g_1g_2a}(T^a)_{BA}
\\
\times
\int\!\frac{d\mathbf{p}}{(2\pi)^3}\int\!\frac{d\mathbf{q}}{(2\pi)^3}
\frac{\bar\Psi^0_{D_{bc}}({\bf p})}
{\sqrt{\frac{\epsilon_c(p)}{m_c}\frac{(\epsilon_c(p)+m_c)}{2m_c}\frac{\epsilon_b(p)}{m_b}\frac{(\epsilon_b(p)+m_b)}{2m_b}}}
\frac{\bar\Psi^0_{\bar D_{\bar b\bar c}}({\bf q})}
{\sqrt{\frac{\epsilon_c(q)}{m_c}\frac{(\epsilon_c(q)+m_c)}{2m_c}\frac{\epsilon_b(q)}{m_b}\frac{(\epsilon_b(q)+m_b)}{2m_b}}}
\\
\times
\mathrm{Tr}\Bigl\{
\Bigl[\frac{\hat v_1-1}2+\hat v_1\frac{\mathbf p^2}{2m_c(\epsilon_c(p)+m_c)}+\frac{\hat p}{2m_c}\Bigr]\Sigma^P(1+\hat v_1)\Bigl[\frac{\hat v_1+1}2+\hat v_1\frac{\mathbf p^2}{2m_b(\epsilon_b(p)+m_b)}-\frac{\hat p}{2m_b}\Bigr]\\
\times \gamma_\beta\frac{m_b-\hat p_1-\hat q_1-\hat q_2}{(p_1+q_1+q_2)-m_b^2}\gamma_\omega \Bigl[\frac{\hat v_2-1}2+\hat v_2\frac{\mathbf q^2}{2m_b(\epsilon_b(q)+m_b)}-\frac{\hat q}{2m_b}\Bigr]\Sigma^Q(1+\hat v_2)\\
\times\Bigl[\frac{\hat v_2+1}2+\hat v_2\frac{\mathbf q^2}{2m_c(\epsilon_c(q)+m_c)}+\frac{\hat q}{2m_c}\Bigr]\gamma_\alpha
\Bigr\}\varepsilon_1^\mu(k_1)\varepsilon_2^\nu(k_2) D^{\alpha\omega}(p_1+q_1)D^{\rho\beta}(k_1+k_2)
\\
\times
\bigl(g_{\mu\nu}(k_2-k_1)_\rho-g_{\nu\rho}(k_1+2k_2)_\mu+g_{\mu\rho}(2k_1+k_2)_\nu\bigr).
\end{gathered}
\end{equation}

The production amplitude~\eqref{eq:m-main} and vertex
functions~\eqref{eq:gammas} contain relative momenta $p$ and $q$ in exact form. In order to take
into account relativistic corrections of second order in $p$ and $q$ we expand all
inverse denominators of the quark and gluon propagators as follows:
\begin{equation}
\begin{gathered}
\label{eq:exps}
\frac{1}{(p_{1,2}+q_{1,2})^2}=\frac{1}{s\,\eta_{1,2}^2}\Bigl[
1 \mp \frac{2(pQ+qP)}{s\,\eta_{1,2}}-\frac{p^2+2pq+q^2}{s\,\eta_{1,2}^2} + \ldots \Bigr], \\
\frac{1}{(p_1+q_1+q_2)^2-m_b^2} =
\frac{1}{Z_1} \Bigl[ 1 - \frac{2pQ+p^2}{Z_1}+\frac{4(pQ)^2}{Z_1^2}  + \ldots \Bigr], \\
\frac{1}{(k_2-q_1)^2-m_c^2} = \frac{1}{Z_2} \Bigl[ 1 + \frac{2k_2q-q^2}{Z_2}+\frac{4(k_2Q)^2}{Z_2^2}  + \ldots \Bigr],
\end{gathered}
\end{equation}
where $Z_1 = s\,\eta_1+\eta_2^2 M^2 - m_b^2$ and $Z_2 = t\,\eta_1+\eta_1\eta_2 M^2 - m_c^2$. The amplitude~\eqref{eq:m-main} contains
16 different denominators to be expanded in the manner of Eq.~\eqref{eq:exps}. Neglecting the bound state corrections, we find
that an expansion of denominators takes one of the following form: $s\,\eta_{1,2}$, $s\,\eta_{1,2}^2$, $\eta_{1,2}(M^2-t)$ or
$\eta_{1,2}(M^2-s-t)$. Then, taking into account kinematical restrictions on $s$ and $t$
\be
4M^2\le s,\quad
\left| t+\frac{s}{2}-M^2 \right|\le\frac{s}{2}\sqrt{1-\frac{4M^2}{s}},
\ee
and nonrelativistic estimate
\hbox{$\eta_1 \approx m_c/(m_c+m_b) \approx 1/4$} for $(bc)$ diquarks, we conclude that expansion parameters in~\eqref{eq:exps}
are at least as small as $4p^2/M^2$ and $4q^2/M^2$.
Preserving in the expanded amplitude terms up to second order both in relative momenta $p$ and $q$,
we can perform the angular integration using the following relations for $\mathcal S$-wave diquarks:
\bdis
\int\!\frac{\Psi^{\mathcal S}_0(\mathbf p)}{\sqrt{ \frac{\epsilon_c(p)}{m_c} \frac{\epsilon_c(p)+m_c}{2m_c}
\frac{\epsilon_b(p)}{m_b} \frac{\epsilon_b(p)+m_b}{2m_b} }}\frac{d\mathbf p}{(2\pi)^3}=
\frac{1}{\sqrt{2}\,\pi}\int\limits_0^\infty\!\frac{p^2R_\mathcal S(p)}{\sqrt{ \frac{\epsilon_c(p)}{m_c}
\frac{\epsilon_c(p)+m_c}{2m_c} \frac{\epsilon_b(p)}{m_b} \frac{\epsilon_b(p)+m_b}{2m_b} }}dp,
\edis
\be
\int\! \frac{p_\mu p_\nu \, \Psi^{\mathcal S}_0(\mathbf p)}{\sqrt{ \frac{\epsilon_c(p)}{m_c}
\frac{\epsilon_c(p)+m_c}{2m_c} \frac{\epsilon_b(p)}{m_b} \frac{\epsilon_b(p)+m_b}{2m_b} }}\frac{d\mathbf p}{(2\pi)^3}=-\frac{g_{\mu\nu}-{v_1}_\mu{v_1}_\nu}{3\sqrt2\,\pi}
\int\limits_0^\infty\!\frac{p^4R_\mathcal S(p)}{\sqrt{ \frac{\epsilon_c(p)}{m_c} \frac{\epsilon_c(p)+m_c}{2m_c}
\frac{\epsilon_b(p)}{m_b} \frac{\epsilon_b(p)+m_b}{2m_b} }}dp,
\ee
where $R_\mathcal S(p)$ is the radial wave function.

In order to calculate the cross section we have to sum the squared modulus of the amplitude over final particle polarizations
in the case of pair axial-vector diquark production and average it over polarizations of initial gluons
using the following relations:
\be
\sum_{\lambda}\varepsilon_{P}^\mu \, {\varepsilon_{P}^\ast}^\nu = v_1^\mu v_1^\nu-g^{\mu\nu},\quad
\sum_{\lambda}\varepsilon_{Q}^\mu \, {\varepsilon_{Q}^\ast}^\nu = v_2^\mu v_2^\nu-g^{\mu\nu},\quad
\sum_{\lambda}\varepsilon_{1,2}^\mu \, \varepsilon_{1,2}^{\ast\;\nu}=\frac{
k_1^\mu k_2^\nu+k_1^\nu k_2^\mu}{k_1\cdot k_2}-g^{\mu\nu}.
\ee
Then we also average it over colors of initial gluons and sum over diquark color indices $A$ and $B$.
Finally, we obtain the following expression for the differential cross section of pair double heavy diquark production:
\bga
\label{eq:cs}
d\sigma[gg\to D_{bc} + \bar D_{\bar b \bar c}](s,t)=\frac{\pi M^2 \alpha_s^4}{65\,536\,s^2}|\tilde R(0)|^4 \times \\
 \bigl[ F^{(1)}(s,t) -4(\omega_{01}+\omega_{10}-\omega_{11}) F^{(1)}(s,t)
-4{m_c^{-1}m_b^{-1}}(m_c^2\omega_{\frac12\frac32} + m_b^2\omega_{\frac32\frac12}) F^{(1)}(s,t) \\
+6(\omega_{01}+\omega_{10})^2 F^{(1)}(s,t) + \omega_{\frac12\frac12}(1-3\omega_{01}-3\omega_{10})F^{(2)}(s,t)
+\omega_{\frac12\frac12}^2 F^{(3)}(s,t) \bigr],
\ega
where the parameter $\tilde R(0)$ in~\eqref{eq:cs} has the following form
\bga
\label{eq:r0}
\tilde R(0) = \sqrt\frac{2}{\pi}\int_0^\infty \!\!\!\sqrt{\frac{(\epsilon_c(p)+m_c)(\epsilon_b(p)+m_b)}{2\epsilon_c(p) \, 2\epsilon_b(p)}} R(p)p^2\, dp.
\ega
It represents the relativistic generalization of the value of wave function at the origin $R(0)$. The relativistic parameters $\omega_{nk}$
are expressed through momentum integrals with the double heavy diquark radial wave function $R(p)$:
\bga
\label{eq:is}
I_{nk} = \int_0^{m_c} \!\!\! p^2 R(p)\sqrt{\frac{(\epsilon_c(p)+m_c)(\epsilon_b(p)+m_b)}{2\epsilon_c(p) \, 2\epsilon_b(p)}}
\biggl( \frac{\epsilon_c(p)-m_c}{\epsilon_c(p)+m_c} \biggr)^n
\biggl( \frac{\epsilon_b(p)-m_b}{\epsilon_b(p)+m_b} \biggr)^k
dp,
\\
\omega_{nk} = \sqrt\frac{2}{\pi}\frac{I_{nk}}{\tilde R(0)}.
\ega
In contrast to our previous work \cite{mt2014} there are terms in~\eqref{eq:cs} which contain relativistic parameters $\omega_{nk}$
with fractional indices. They appear if we preserve the symmetry of cross section~\eqref{eq:cs} in quark masses $m_c$ and $m_b$.
The auxiliary functions $F^{(i)}(s,t)$ contain nonrelativistic contribution and relativistic corrections to the cross section connected
with the relative motion of heavy quarks. Their exact analytical expressions are extremely lengthy in the case of diquarks of different
flavors $b$ and $c$, so we present them in Appendix A only in the case of $(cc)$ diquarks.

\section{Numerical results and discussion}

The quasipotential wave functions of double heavy diquarks are obtained by numerical solution of the Schr\"odinger equation with
effective relativistic Hamiltonian based on the QCD generalization of the Breit potential completed by scalar and vector exchange confinement terms,
as it is described in details in our previous works~\cite{mt2012,mt2014}. We present the values of diquark masses and relativistic parameters~\eqref{eq:r0}, \eqref{eq:is} in Table~\ref{tbl0}. Numerical masses of charmonium and $B_c$ mesons obtained in our model are in good agreement with existing
experimental data (the difference is less than 1\%)~\cite{ebert,mt2012,mt2014}. Analogously, the masses for $(bc)$ and $(cc)$ double heavy diquarks from Table~\ref{tbl0} coincide with the estimates made in other approaches~\cite{baryon,gershtein,ebert3,ebert07}. Note that our definition~\eqref{eq:is}
of relativistic integrals $I_{nk}$ contains a cutoff at the value of $c$-quark mass $\Lambda = m_c$. Although the integrals~\eqref{eq:is} are convergent,
there are some uncertainties in their calculation related with the determination of the wave function in the region of relativistic momenta $p\gtrsim m_c$
in our model.

\begin{table*}[t]
\caption{\label{tbl0}Numerical values of parameters describing double heavy $(cc)$ and $(bc)$ diquarks}
\bigskip
\begin{ruledtabular}
\begin{tabular}{|l|c|c|c|c|c|c|c|c|c|}
\hline
Diquark & $n^{2S+1}L_J$ & $M$, & $\tilde R(0)$, & $\omega_{10}$ &$\omega_{01}$ & $\omega_{\frac12\frac12}$   &
$\omega_{11}$ &  $\omega_{\frac12\frac32}$ &  $\omega_{\frac32\frac12}$ \\
state  &     &   GeV  &   GeV$^{3/2}$  &    &     &    &    &  &  \\    \hline
$SD_{bc}$	& $1^1S_0$	& 6.517 & 0.50 & 0.0383 & 0.0045 & 0.0131  & 0.00039    & 0.00014 & 0.0011 \\ \hline
$AVD_{bc}$  & $1^3S_1$  & 6.526 & 0.48 & 0.0384 & 0.0045 & 0.0132  & 0.00038    & 0.00013 & 0.0011 \\  \hline
$AVD_{cc} $ & $1^3S_1$	& 3.224 & 0.38 &\multicolumn{3}{c|}{0.0323}&\multicolumn{3}{c|}{0.0023}   \\  \hline
\end{tabular}
\end{ruledtabular}
\end{table*}

\begin{table*}[t]
\caption{\label{tbl1}The cross section values of pair double heavy diquark production.}
\begin{ruledtabular}
\begin{tabular}{|c|l|l|l|l|l|}
Energy $\sqrt S$ &
Diquark pair & \multicolumn{2}{c|}{CTEQ5L}  & \multicolumn{2}{c|}{CTEQ6L1}
\\
& & \multicolumn{1}{l}{$\sigma_{nonrel}$,~nb} & $\sigma_{rel}$,~nb & \multicolumn{1}{l}{$\sigma_{nonrel}$,~nb} & $\sigma_{rel}$,~nb \\ \hline
$\sqrt S=7$~TeV &
\begin{tabular}{l} $SD_{bc}+S\bar D_{\bar b\bar c}$ \\ $AVD_{bc}+AV\bar D_{\bar b\bar c}$ \\ $AVD_{cc}+AV\bar D_{\bar c\bar c}$
\end{tabular} &
\multicolumn{1}{l}{
\begin{tabular}{c} 0.063 \\ 0.25 \\ 1.39 \end{tabular}} &
\begin{tabular}{c} 0.018 \\ 0.053 \\ 0.28 \end{tabular} &
\multicolumn{1}{l}{
\begin{tabular}{c} 0.057 \\ 0.23 \\ 1.07 \end{tabular}} &
\begin{tabular}{c} 0.016 \\ 0.049 \\ 0.22 \end{tabular} \\ \hline
$\sqrt S=14$~TeV &
\begin{tabular}{l} $SD_{bc}+S\bar D_{\bar b\bar c}$ \\ $AVD_{bc}+AV\bar D_{\bar b\bar c}$ \\ $AVD_{cc}+AV\bar D_{\bar c\bar c}$
\end{tabular} &
\multicolumn{1}{l}{
\;\begin{tabular}{c} 0.14 \\ 0.55 \\ 2.51 \end{tabular}} &
\begin{tabular}{c} 0.039 \\ 0.12 \\ 0.51 \end{tabular} &
\multicolumn{1}{l}{
\;\begin{tabular}{c} 0.12 \\ 0.48 \\ 1.94 \end{tabular}} &
\begin{tabular}{c} 0.034 \\ 0.10 \\ 0.40 \end{tabular}
\end{tabular}
\end{ruledtabular}
\end{table*}

\begin{table*}[t]
\caption{\label{tbl2}Double heavy diquark production cross sections corresponding to the rapidity range  $2<y_{P,Q}<4.5$}
\begin{ruledtabular}
\begin{tabular}{|c|l|l|l|l|l|}
Energy $\sqrt S$ &
Diquark pair & \multicolumn{2}{c|}{CTEQ5L}  & \multicolumn{2}{c|}{CTEQ6L1}
\\
& & \multicolumn{1}{l}{$\sigma_{nonrel}$,~nb} & $\sigma_{rel}$,~nb & \multicolumn{1}{l}{$\sigma_{nonrel}$,~nb} & $\sigma_{rel}$,~nb \\ \hline
$\sqrt S=7$~TeV &
\begin{tabular}{l} $SD_{bc}+S\bar D_{\bar b\bar c}$ \\ $AVD_{bc}+AV\bar D_{\bar b\bar c}$ \\ $AVD_{cc}+AV\bar D_{\bar c\bar c}$
\end{tabular} &
\multicolumn{1}{l}{
\begin{tabular}{c} 0.010 \\ 0.032 \\ 0.19 \end{tabular}} &
\begin{tabular}{c} 0.003 \\ 0.007 \\ 0.038 \end{tabular} &
\multicolumn{1}{l}{
\begin{tabular}{c} 0.009 \\ 0.029 \\ 0.14 \end{tabular}} &
\begin{tabular}{c} 0.003 \\ 0.006 \\ 0.029 \end{tabular} \\ \hline
$\sqrt S=14$~TeV &
\begin{tabular}{l} $SD_{bc}+S\bar D_{\bar b\bar c}$ \\ $AVD_{bc}+AV\bar D_{\bar b\bar c}$ \\ $AVD_{cc}+AV\bar D_{\bar c\bar c}$
\end{tabular} &
\multicolumn{1}{l}{
\begin{tabular}{c} 0.024 \\ 0.076 \\ 0.35 \end{tabular}} &
\begin{tabular}{c} 0.007 \\ 0.016 \\ 0.072 \end{tabular} &
\multicolumn{1}{l}{
\begin{tabular}{c} 0.020 \\ 0.066 \\ 0.25 \end{tabular}} &
\begin{tabular}{c} 0.006 \\ 0.014 \\ 0.053 \end{tabular}
\end{tabular}
\end{ruledtabular}
\end{table*}

The numerical results for the total cross section of pair double heavy diquark production corresponding to the LHC relative
energies $\sqrt S=7$ and 14~TeV are presented in Table~\ref{tbl1}. The integration in~\eqref{eq:cs-plus-x} is performed with
partonic distribution functions from CTEQ5L and CTEQ6L1 sets \cite{CTEQ5L,CTEQ6L1}. The renormalization and factorization scales
are set equal to transverse mass $\mu=m_T=\sqrt{M^2+P_T^2}$. The leading order result for strong coupling constant $\alpha_s(\mu)$
with initial value $\alpha_s(\mu=M_Z)=0.118$ is used. On second stage
the diquark nucleus can join with high probability a light quark and form double heavy baryon.
In nonrelativistic limit all parameters $\omega_{nk}$ are equal zero and only $F^{(1)}(s,t)$ term survives in square brackets of~\eqref{eq:cs}.
Then, replacing $\tilde R(0)$ by nonrelativistic value of radial wave function at the origin $R(0)=\sqrt{2/\pi}\int p^2R(p)\,dp$ and assuming that
diquark mass is equal to the sum of masses of constituent quarks $M_0=m_b + m_c$, we obtain our nonrelativistic prediction for pair double heavy
diquark production cross section presented in third and fifth columns of Tables~\ref{tbl1} and~\ref{tbl2}. In our model we obtain the following
nonrelativistic values $R(0)=0.67$~GeV$^{3/2}$ and $R(0)=0.53$~GeV$^{3/2}$  for $(bc)$ and $(cc)$ diquarks respectively, which lie close to results $R(0)=0.73$~GeV$^{3/2}$ and $R(0)=0.53$~GeV$^{3/2}$ from~\cite{baryon,gershtein,ebert3}. In order to obtain the cross sections for pair axial-vector $(cc)$
diquark production we replace $m_b\to m_c$  in all expressions and multiply the amplitude by an additional factor $1/4$ ($1/16$ in the cross section)
according to the Pauli exclusion principle.

As it follows from the results presented in Tables~\ref{tbl1} and~\ref{tbl2}, relativistic effects almost five times decrease the values
of $(bc)$ and $(cc)$ pair double heavy diquark production cross sections. The main role in such decrease plays the difference between
relativistic parameter $\tilde R(0)$ and nonrelativistic one $R(0)$. $\tilde R(0)$ ($R(0)$) enters the corresponding cross section in fourth degree,
so that even small modification of this parameter caused by relativistic corrections in the Breit potential leads to a substantial change
in the cross section. For example, in the case of axial-vector $(bc)$ diquark $\tilde R(0)$ is only 25\% smaller than its nonrelativistic value, but
this difference results in more than three times decrease of the cross section value. The bound state effects connected with the non-zero diquark
bound state energy $W = M- m_c -m_b\ne0$ bring an additional 30\% decrease. Finally, the relativistic corrections originating from the expansion of
the production amplitude increase the cross section value by 10--20\%, that is insufficient to compensate the large negative contributions from
the first two sources.

\begin{figure}[t]
\center
\includegraphics{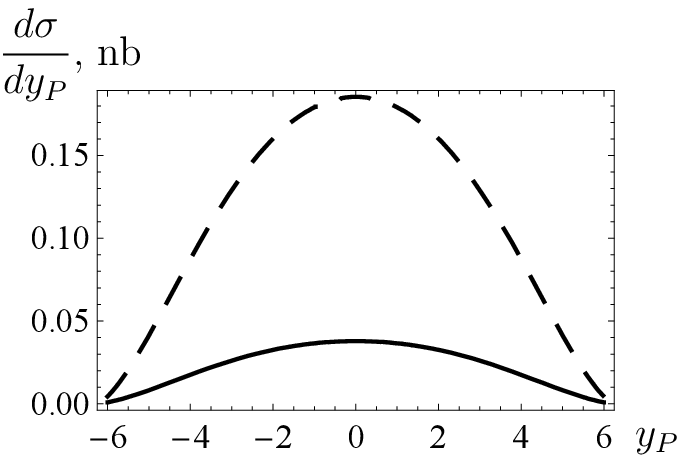}\hfill\includegraphics{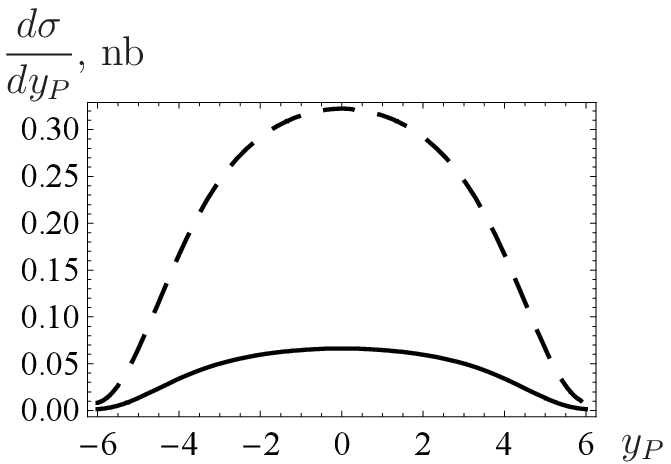}
\caption{The differential cross sections for $pp\to D\bar D+X$
at $\sqrt{S}=7$~TeV (left) and $\sqrt{S}=14$~TeV (right)
as functions of rapidity~$y_P$.
Solid and dashed curves represent total and nonrelativistic
results respectively.}
\label{fig3:dsydd}
\end{figure}

\begin{figure}[t]
\center
\includegraphics{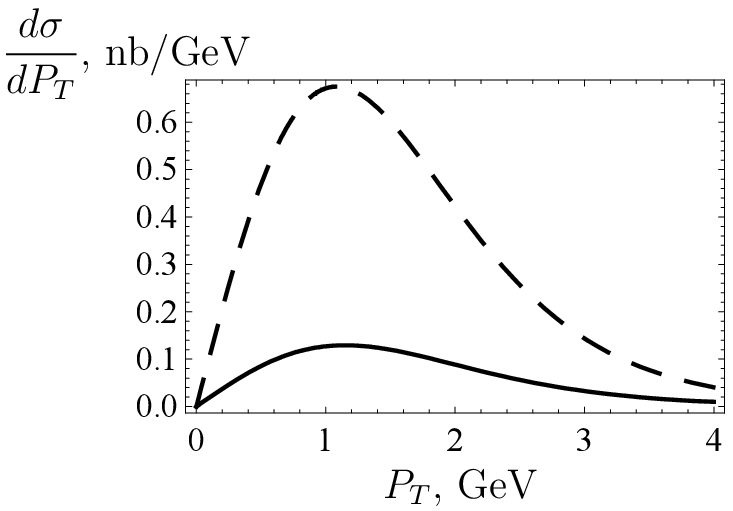}\hfill\includegraphics{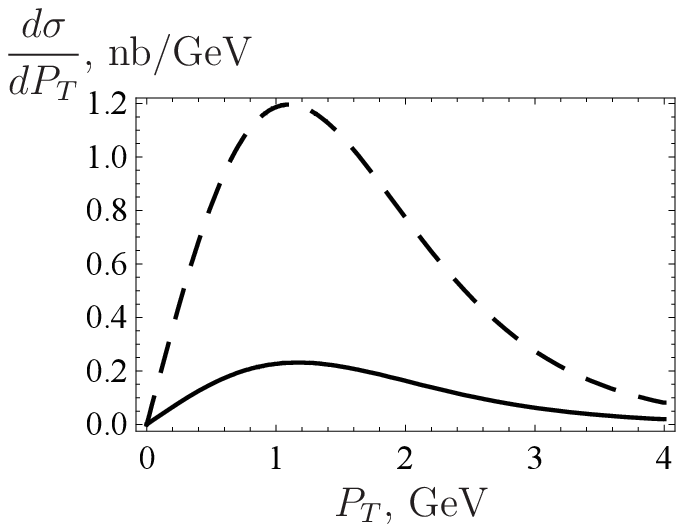}
\caption{The differential cross sections for $pp\to D\bar D+X$
at $\sqrt{S}=7$~TeV (left) and $\sqrt{S}=14$~TeV (right)
as functions of transverse momentum $P_T$ of diquark pair integrated over the
rapidity. Solid and dashed curves represent total and nonrelativistic
results respectively.}
\label{fig4:ptdd}
\end{figure}

\begin{figure}[t]
\center
\includegraphics[scale=0.9]{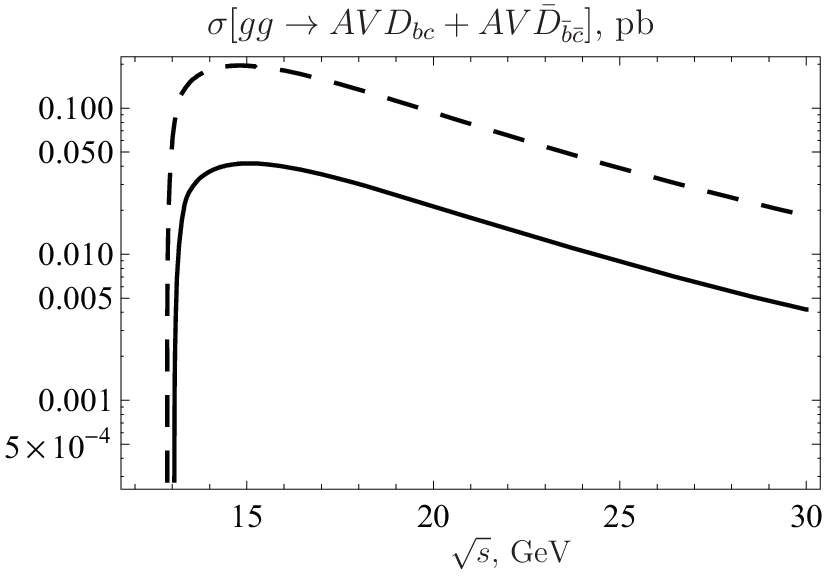}\hfill\includegraphics[scale=0.8]{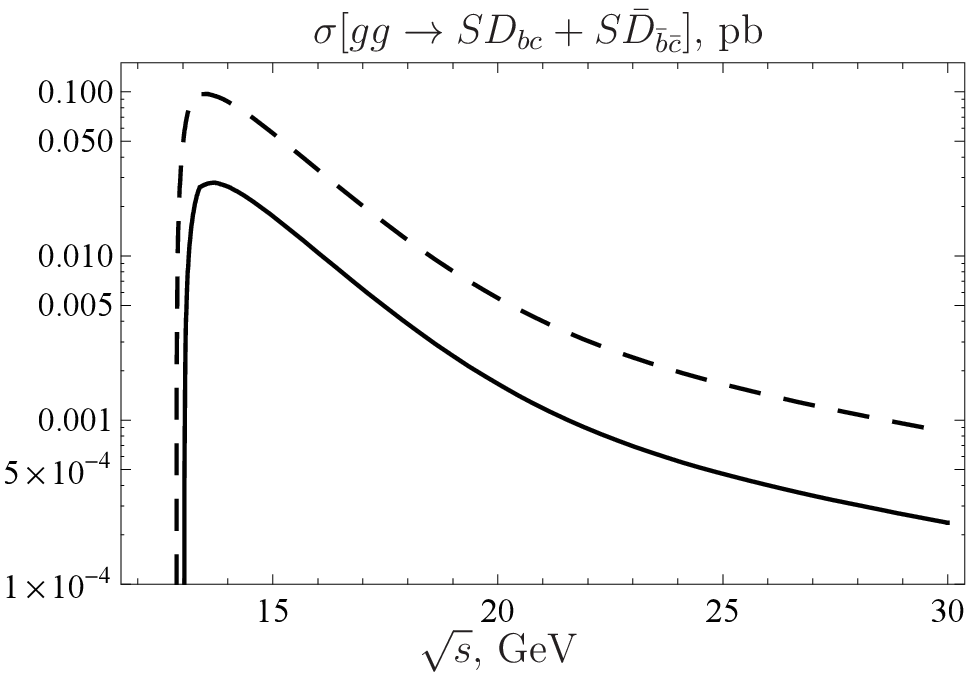}
\caption{Cross section of diquark-antidiquark gluonic production as a
function of their invariant mass.
Solid and dashed curves represent total and nonrelativistic
results respectively.}
\label{fig5:sqrts}
\end{figure}

In Fig.~\ref{fig3:dsydd} we present the results of our calculation of the differential cross section in terms of the
rapidity $y_P=\frac{1}{2}\ln\frac{P_0+P_{\parallel}}{P_0-P_{\parallel}}$. The rapidities of outcoming
diquarks with momenta $P$ and $Q$ can be obtained in the form:
\be
\label{eq:raps}
y_{P,Q}=\frac12\ln\frac{x_1}{x_2}\pm\frac12\ln\left[ \frac{s}{M^2-t} -1 \right].
\ee
The differential cross section $d\sigma/dy_P$ shown in Fig.~\ref{fig3:dsydd} can be important for a comparison
with forthcoming experimental data. It is clear from this plot that relativistic effects strongly
influence on the rapidity distribution of the final diquarks.
In the LHCb experiment~\cite{LHCb} the rapidity lies in the range \hbox{$2<y_{P,Q}<4.5$},
so we should integrate the differential cross section~\eqref{eq:cs-plus-x} over rapidities from
such interval in order to obtain the value corresponding to the experiment at the LHCb detector.
These results are presented in Table~\ref{tbl2}. We show in Fig.~\ref{fig4:ptdd} the distribution over transverse momentum of the
diquarks integrated over all rapidities at $\sqrt{S}=7$~TeV.
It can be seen in Fig.~\ref{fig4:ptdd} that the account of relativistic corrections leads to the
ratio of relativistic and nonrelativistic cross sections $\sigma_{rel}/\sigma_{nr}\approx 0.2$ near the
peak. This trend remains unchanged in the region of high transverse momenta.
In order to have more complete concept about production processes we show in Fig.~\ref{fig5:sqrts}
the cross section of double diquark production in the gluonic subprocess as a function of
its invariant mass.
As it follows from Figs.~\ref{fig4:ptdd}-\ref{fig5:sqrts} that a typical $p_T$ momentum is of order 1.2 GeV
and total typical momenta of diquarks and anti-diquarks is more than 2.5 GeV.
The most part of the pair $(bc)$ diquark production cross section
is accumulated in that region of $\sqrt{s}$ which corresponds to large momenta ${|\bf P}|\geq 2.5$ GeV: $70 \%$ for pair scalar
diquarks and $85\%$ for pair axial vector diquarks.
But the probability $|\Psi_0^{\cal S}({\bf p})|^2$ to find quarks with
relativistic relative momentum $p\geq 1.5$ GeV is strongly suppressed (we use a cutoff for momentum
integrals in (23) at $m_c=1.55$ GeV). This follows from the obtained relativistic wave functions in
our model which have maximum values at $p,q\sim 0.4$ GeV. So, we could expect that
four heavy quarks and anti-quarks are not sufficiently close in the phase space and
rescattering effects between heavy quarks and anti-quarks are not large, but they should be investigated
additionally.

Let us estimate the total theoretical uncertainty of obtained results. The first and main source of the uncertainty is connected with
the relativistic parameter $\tilde R(0)$, which enters the cross section in fourth degree and defines the order of magnitude of final
result. The accuracy of this parameter depends directly on the error in the determination of relativistic quasipotential wave function in
our model, which we estimate in 10\%. Of course, this estimate is a very approximate one but it can be justified by the better than one percent
accuracy of the calculation of charmonium mass spectrum. Then, we estimate the error in the cross section from this source as not exceeding 40\%.
The next source of uncertainty deals with the corrections of fourth and highest order, which are truncated in our amplitude
expansions~\eqref{eq:exps}. As it was mentioned before, the corrections of second order give 10--20\% contribution to the cross section
value, so we suppose that 20\% will be the reasonable estimate for this error. The contribution of the next-to-leading order in strong coupling
constant $\alpha_s$ is difficult to estimate. It depends significantly on the structure of the Feynman amplitude and has to be calculated independently.
For example, it is known that such corrections lead to significantly increasing factor $K=1.6\div 1.9$ to the cross section
of pair charmonium production in $e^+e^-$ annihilation~\cite{chao,eenlo}. On the other hand, recently it was found that NLO $\alpha_s$ contribution
to the cross section of pair $J/\psi$ production in $pp$-collision for the LHCb rapidity range amounts the value of order 10\%~\cite{jpsinlo}.
So, we assume that similar contribution occurs in pair diquark production in $pp$-collision. Finally, there is one additional uncertainty
connected with the accuracy of partonic distribution functions, which was estimated to be 15\% in~\cite{mt2012}. Then, adding all the
mentioned uncertainties in quadrature, we obtain the total error in 48\% for our results.

\acknowledgments

The authors are grateful to D.~Ebert, R.N.~Faustov and V.O.~Galkin for useful discussions.
The work is supported partially by the Ministry of Education and Science of Russian Federation
(government order for Samara State U. No. 2.870.2011) and Dynasty Foundation.

\appendix
\section{The coefficients $F^{(i)}$ entering the differential cross section~(\ref{eq:cs}) for pair axial-vector
diquark $(cc)$ production.}
\label{app:fis}

Exact analytical expressions for functions $F^{(i)}(s,t)$ in~\eqref{eq:cs} are extremely lengthy for heavy quarks of
different flavor, so we present here only their analytical expressions in the case of pair axial-vector diquark $(cc)$ production.
In these expressions we take into account linear effects in the bound state energy W of heavy quarks and introduce the notation
$M_c=2m_c$. Bound state effects are taken into account in numerical results from Tables~\ref{tbl1} and \ref{tbl2}.
\begin{equation}
\label{eq:fist}
F^{(i)}(s,t)=F^{(i)}_0(s,t)+W\,F^{(i)}_1(s,t),
\end{equation}
{
\allowdisplaybreaks
\begin{align*}
F&^{(1)}_0[AVD_{cc}+AV\bar D_{\bar c\bar c}](s,t) = \frac{524288}{27 M_c^4 s^6 \left(M_c^2-t\right)^4 \left(M_c^2-s-t\right)^4} \\ &
\!\times\!
\bigl[
27648 M_c^{24}\!-\!72 M_c^{22} (1595 s\!+\!4596 t)\!+\!3 M_c^{20} (67687 s^2\!+\!437088 s t\!+\!605232 t^2)\\&
\!-\!8 M_c^{18} (28007 s^3\!+\!278328 s^2 t\!+\!849501 s t^2\!+\!754920 t^3)\!+\!4 M_c^{16} (48546 s^4\!+\!575480 s^3 t\\&
\!+\!2731629 s^2 t^2\!+\!5276664 s t^3\!+\!3390660 t^4)\!-\!2 M_c^{14} (66854 s^5\!+\!867710 s^4 t\!+\!5237453 s^3 t^2\\&
\!+\!15810492 s^2 t^3\!+\!21825720 s t^4\!+\!10831968 t^5)\!+\!M_c^{12} (64025 s^6\!+\!980113 s^5 t\!+\!6934011 s^4 t^2\\&
\!+\!27679700 s^3 t^3\!+\!59798910 s^2 t^4\!+\!63129024 s t^5\!+\!25238304 t^6)\!-\!2 M_c^{10} (9796 s^7\\&
\!+\!190998 s^6 t\!+\!1629993 s^5 t^2\!+\!8003124 s^4 t^3\!+\!23392115 s^3 t^4\!+\!38627220 s^2 t^5\!+\!32576040 s t^6\\&
\!+\!10803456 t^7)\!+\!t^2 (s\!+\!t)^2 (8 s^8\!+\!25 s^7 t\!+\!2536 s^6 t^2\!+\!21366 s^5 t^3\!+\!78759 s^4 t^4\!+\!157896 s^3 t^5\\&
\!+\!179640 s^2 t^6\!+\!108864 s t^7\!+\!27216 t^8)\!-\!2 M_c^2 t (s\!+\!t)^2 (16 s^8\!+\!243 s^7 t\!+\!5526 s^6 t^2\\&
\!+\!49040 s^5 t^3\!+\!215626 s^4 t^4\!+\!530597 s^3 t^5\!+\!741924 s^2 t^6\!+\!546660 s t^7\!+\!163296 t^8)\!+\!M_c^8 \\&
\!\times\! (4006 s^8\!+\!94606 s^7 t\!+\!1029199 s^6 t^2\!+\!6247798 s^5 t^3\!+\!23171033 s^4 t^4\!+\!52444016 s^3 t^5\\&
\!+\!69078684 s^2 t^6\!+\!47988288 s t^7\!+\!13491360 t^8)\!-\!2 M_c^6 (322 s^9\!+\!7064 s^8 t\!+\!99306 s^7 t^2\\&
\!+\!779460 s^6 t^3\!+\!3657884 s^5 t^4\!+\!10718238 s^4 t^5\!+\!19496435 s^3 t^6\!+\!21114948 s^2 t^7\\&
\!+\!12361788 s t^8\!+\!2995920 t^9)\!+\!M_c^4 (68 s^{10}\!+\!1153 s^9 t\!+\!19692 s^8 t^2\!+\!217805 s^7 t^3\\&
\!+\!1362129 s^6 t^4\!+\!5166549 s^5 t^5\!+\!12342213 s^4 t^6\!+\!18546596 s^3 t^7\\&
\!+\!16897269 s^2 t^8\!+\!8485920 s t^9\!+\!1796688 t^{10})
\bigr],
\addtocounter{equation}{1}
\tag{\theequation}
\end{align*}
\begin{align*}
F&^{(2)}_0[AVD_{cc}+AV\bar D_{\bar c\bar c}](s,t) = -\frac{524288}{81 M_c^4 s^8 \left(M_c^2-t\right)^5 \left(M_c^2-s-t\right)^5} \\ &
\!\times\!
\bigl[
1741824 M_c^{32}\!-\!1728 M_c^{30} (5897 s\!+\!16128 t)\!+\!144 M_c^{28} (189641 s^2\!+\!1087056 s t\!+\!1451520 t^2)\\&
\!-\!144 M_c^{26} (322301 s^3\!+\!2802300 s^2 t\!+\!7789176 s t^2\!+\!6773760 t^3)\!+\!M_c^{24} (57554074 s^4\\&
\!+\!656045232 s^3 t\!+\!2763480672 s^2 t^2\!+\!4972983552 s t^3\!+\!3170119680 t^4)\!-\!2 M_c^{22} (27433899 s^5\\&
\!+\!384167638 s^4 t\!+\!2133779508 s^3 t^2\!+\!5813307936 s^2 t^3\!+\!7628766624 s t^4\!+\!3804143616 t^5)\\&
\!+\!6 M_c^{20} (6821962 s^6\!+\!113839005 s^5 t\!+\!781665683 s^4 t^2\!+\!2819646560 s^3 t^3\!+\!5596217064 s^2 t^4\\&
\!+\!5718898944 s t^5\!+\!2324754432 t^6)\!-\!2 M_c^{18} (12207465 s^7\!+\!234642355 s^6 t\!+\!1930593059 s^5 t^2\\&
\!+\!8652159088 s^4 t^3\!+\!22804443900 s^3 t^4\!+\!35210657664 s^2 t^5\!+\!29218299264 s t^6\!+\!9963233280 t^7)\\&
\!+\!M_c^{16} (11661347 s^8\!+\!251850550 s^7 t\!+\!2421795360 s^6 t^2\!+\!13072181288 s^5 t^3\!+\!42975298740 s^4 t^4\\&
\!+\!88311607776 s^3 t^5\!+\!110619734400 s^2 t^6\!+\!76740162048 s t^7\!+\!22417274880 t^8)\\&
\!-\!M_c^{14} (4223561 s^9\!+\!104555317 s^8 t\!+\!1160158397 s^7 t^2\!+\!7399876560 s^6 t^3\!+\!29431644460 s^5 t^4\\&
\!+\!75678177864 s^4 t^5\!+\!126369906768 s^3 t^6\!+\!132222824064 s^2 t^7\!+\!78350462784 s t^8\!+\!19926466560 t^9)\\&
\!+\!4 t^3 (s\!+\!t)^3 (88 s^{10}\!+\!291 s^9 t\!+\!30322 s^8 t^2\!+\!297142 s^7 t^3\!+\!1336999 s^6 t^4\!+\!3529608 s^5 t^5\\&
\!+\!5988728 s^4 t^6\!+\!6737472 s^3 t^7\!+\!4950288 s^2 t^8\!+\!2177280 s t^9\!+\!435456 t^{10})\!+\!M_c^{12} (1077011 s^{10}\\&
\!+\!31978294 s^9 t\!+\!416887472 s^8 t^2\!+\!3125057808 s^7 t^3\!+\!14810788148 s^6 t^4\!+\!46251513620 s^5 t^5\\&
\!+\!96893779668 s^4 t^6\!+\!135324576576 s^3 t^7\!+\!120857286384 s^2 t^8\!+\!62193643776 s t^9\!+\!13948526592 t^{10})\\&
\!-\!M_c^2 t^2 (s\!+\!t)^2 (1504 s^{11}\!+\!23728 s^{10} t\!+\!685939 s^9 t^2\!+\!7528052 s^8 t^3\!+\!42573677 s^7 t^4\\&
\!+\!146340250 s^6 t^5\!+\!329945926 s^5 t^6\!+\!504952416 s^4 t^7\!+\!526855224 s^3 t^8\!+\!362855808 s^2 t^9\\&
\!+\!149506560 s t^{10}\!+\!27869184 t^{11})\!-\!M_c^{10} (189099 s^{11}\!+\!6711431 s^{10} t\!+\!106896474 s^9 t^2\\&
\!+\!958886210 s^8 t^3\!+\!5412164759 s^7 t^4\!+\!20272621428 s^6 t^5\!+\!51752902204 s^5 t^6\!+\!90884023920 s^4 t^7\\&
\!+\!108459347328 s^3 t^8\!+\!84070738368 s^2 t^9\!+\!38069792640 s t^{10}\!+\!7608287232 t^{11})\!+\!M_c^8 (23696 s^{12}\\&
\!+\!902460 s^{11} t\!+\!18000829 s^{10} t^2\!+\!201542412 s^9 t^3\!+\!1383434183 s^8 t^4\!+\!6237258490 s^7 t^5\\&
\!+\!19205490660 s^6 t^6\!+\!41225350568 s^5 t^7\!+\!61985442066 s^4 t^8\!+\!64263674160 s^3 t^9\!+\!43813561824 s^2 t^{10}\\&
\!+\!17647352064 s t^{11}\!+\!3170119680 t^{12})\!+\!2 M_c^4 t (1648 s^{13}\!+\!36454 s^{12} t\!+\!814068 s^{11} t^2\!+\!10491584 s^{10} t^3\\&
\!+\!77792679 s^9 t^4\!+\!366446361 s^8 t^5\!+\!1170001946 s^7 t^6\!+\!2627856120 s^6 t^7\!+\!4228552375 s^5 t^8\\&
\!+\!4881077725 s^4 t^9\!+\!3965125248 s^3 t^{10}\!+\!2157018696 s^2 t^{11}\!+\!705148416 s t^{12}\!+\!104509440 t^{13})\\&
\!-\!M_c^6 (1664 s^{13}\!+\!72088 s^{12} t\!+\!1733298 s^{11} t^2\!+\!25982195 s^{10} t^3\!+\!228949037 s^9 t^4\\&
\!+\!1276267953 s^8 t^5\!+\!4775775315 s^7 t^6\!+\!12428268880 s^6 t^7\!+\!22908511678 s^5 t^8\!+\!29980767340 s^4 t^9\\&
\!+\!27364070472 s^3 t^{10}\!+\!16589102400 s^2 t^{11}\!+\!5996839104 s t^{12}\!+\!975421440 t^{13})
\bigr],
\addtocounter{equation}{1}
\tag{\theequation}
\end{align*}
\begin{align*}
F&^{(3)}_0[AVD_{cc}+AV\bar D_{\bar c\bar c}](s,t) = \frac{262144}{243 M_c^4 s^{10} \left(M_c^2-t\right)^6 \left(M_c^2-s-t\right)^6} \\ &
\!\times\!
\bigl[
222953472 M_c^{40}\!-\!27869184 M_c^{38} (59 s\!+\!160 t)\!+\!3456 M_c^{36} (1634233 s^2\!+\!9209088 s t\!+\!12257280 t^2)\\&
\!-\!576 M_c^{34} (21154723 s^3\!+\!183547236 s^2 t\!+\!506435328 s t^2\!+\!441262080 t^3)\!+\!48 M_c^{32} (390910561 s^4\\&
\!+\!4582950624 s^3 t\!+\!19453613688 s^2 t^2\!+\!35059433472 s t^3\!+\!22504366080 t^4)\!-\!4 M_c^{30} (5537521813 s^5\\&
\!+\!81274350792 s^4 t\!+\!466630287984 s^3 t^2\!+\!1292676789888 s^2 t^3\!+\!1712700702720 s t^4\\&
\!+\!864167657472 t^5)\!+\!4 M_c^{28} (5240554941 s^6\!+\!91437358522 s^5 t\!+\!659112431964 s^4 t^2\\&
\!+\!2471543238528 s^3 t^3\!+\!5029243275744 s^2 t^4\!+\!5227645796352 s t^5\!+\!2160419143680 t^6)\\&
\!-\!M_c^{26} (16330544031 s^7\!+\!326792003628 s^6 t\!+\!2815199542892 s^5 t^2\!+\!13284503935680 s^4 t^3\\&
\!+\!36610822146240 s^3 t^4\!+\!58396559171328 s^2 t^5\!+\!49627101855744 s t^6\!+\!17283353149440 t^7)\\&
\!+\!M_c^{24} (10537886063 s^8\!+\!237559033812 s^7 t\!+\!2365887361944 s^6 t^2\!+\!13397155539088 s^5 t^3\\&
\!+\!46551035169840 s^4 t^4\!+\!100601034937344 s^3 t^5\!+\!131082206097024 s^2 t^6\!+\!93716708327424 s t^7\\&
\!+\!28085448867840 t^8)\!-\!2 M_c^{22} (2784626670 s^9\!+\!70737215246 s^8 t\!+\!800148664831 s^7 t^2\\&
\!+\!5268464876020 s^6 t^3\!+\!22039344056906 s^5 t^4\!+\!60145366265904 s^4 t^5\!+\!106205291855712 s^3 t^6\\&
\!+\!116324448634368 s^2 t^7\!+\!71451599192064 s t^8\!+\!18723632578560 t^9)\!+\!M_c^{20} (2378426106 s^{10}\\&
\!+\!68538265551 s^9 t\!+\!877306323899 s^8 t^2\!+\!6598930104580 s^7 t^3\!+\!32238538637660 s^6 t^4\\&
\!+\!106201836358040 s^5 t^5\!+\!237119424205104 s^4 t^6\!+\!352010470800384 s^3 t^7\!+\!330977605825152 s^2 t^8\\&
\!+\!177504964116480 s t^9\!+\!41191991672832 t^{10})\!-\!M_c^{18} (815322191 s^{11}\!+\!26692617042 s^{10} t\\&
\!+\!387726963435 s^9 t^2\!+\!3311677836814 s^8 t^3\!+\!18562148953601 s^7 t^4\!+\!71659299951396 s^6 t^5\\&
\!+\!193566257649996 s^5 t^6\!+\!363728035723392 s^4 t^7\!+\!463561912782144 s^3 t^8\!+\!380261571985152 s^2 t^9\\&
\!+\!180350463541248 s t^{10}\!+\!37447265157120 t^{11})\!+\!4 t^4 (s\!+\!t)^4 (2904 s^{12}\!+\!10131 s^{11} t\!+\!1088812 s^{10} t^2\\&
\!+\!12565370 s^9 t^3\!+\!71302945 s^8 t^4\!+\!257354136 s^7 t^5\!+\!650793576 s^6 t^6\!+\!1191722688 s^5 t^7\\&
\!+\!1576367280 s^4 t^8\!+\!1465413120 s^3 t^9\!+\!906204672 s^2 t^{10}\!+\!334430208 s t^{11}\!+\!55738368 t^{12})\\&
\!+\!M_c^{16} (220505891 s^{12}\!+\!8239519834 s^{11} t\!+\!136412070144 s^{10} t^2\!+\!1324461974452 s^9 t^3\\&
\!+\!8455632770450 s^8 t^4\!+\!37581776673544 s^7 t^5\!+\!119310422197680 s^6 t^6\!+\!271757302270944 s^5 t^7\\&
\!+\!438797604649680 s^4 t^8\!+\!487891081751040 s^3 t^9\!+\!353665494179712 s^2 t^{10}\!+\!149887606063104 s t^{11}\\&
\!+\!28085448867840 t^{12})\!-\!M_c^2 t^3 (s\!+\!t)^3 (52864 s^{13}\!+\!808320 s^{12} t\!+\!29421610 s^{11} t^2\!+\!375769223 s^{10} t^3\\&
\!+\!2557032615 s^9 t^4\!+\!11233867505 s^8 t^5\!+\!34796810259 s^7 t^6\!+\!79006480208 s^6 t^7\!+\!132647858276 s^5 t^8\\&
\!+\!162616078080 s^4 t^9\!+\!141063190272 s^3 t^{10}\!+\!81801031680 s^2 t^{11}\!+\!28398698496 s t^{12}\!+\!4459069440 t^{13})\\&
\!-\!2 M_c^{14} (22481965 s^{13}\!+\!978678070 s^{12} t\!+\!18711980507 s^{11} t^2\!+\!207897944100 s^{10} t^3\\&
\!+\!1512657963970 s^9 t^4\!+\!7680203547784 s^8 t^5\!+\!28156966746706 s^7 t^6\!+\!75568224170616 s^6 t^7\\&
\!+\!148153722926478 s^5 t^8\!+\!208861152514800 s^4 t^9\!+\!205204061269152 s^3 t^{10}\!+\!132824617585920 s^2 t^{11}\\&
\!+\!50738580652032 s t^{12}\!+\!8641676574720 t^{13})\!+\!M_c^4 t^2 (s\!+\!t)^2 (122528 s^{14}\!+\!2904080 s^{13} t\\&
\!+\!82920640 s^{12} t^2\!+\!1181123451 s^{11} t^3\!+\!9594060453 s^{10} t^4\!+\!51055788685 s^9 t^5\!+\!192342235019 s^8 t^6\\&
\!+\!534248481092 s^7 t^7\!+\!1111820399652 s^6 t^8\!+\!1729605262024 s^5 t^9\!+\!1976699347728 s^4 t^{10}\\&
\!+\!1605475058688 s^3 t^{11}\!+\!874662022656 s^2 t^{12}\!+\!285993566208 s t^{13}\!+\!42361159680 t^{14})\\&
\!+\!M_c^{12} (6597472 s^{14}\!+\!338366703 s^{13} t\!+\!7696397325 s^{12} t^2\!+\!99678452662 s^{11} t^3\!+\!833912155244 s^{10} t^4\\&
\!+\!4841060642370 s^9 t^5\!+\!20335964523054 s^8 t^6\!+\!63209010354088 s^7 t^7\!+\!146358554142828 s^6 t^8\\&
\!+\!250907906992024 s^5 t^9\!+\!312764014124112 s^4 t^{10}\!+\!274363712128512 s^3 t^{11}\!+\!159957170755200 s^2 t^{12}\\&
\!+\!55477289484288 s t^{13}\!+\!8641676574720 t^{14})\!-\!2 M_c^6 t (s\!+\!t) (56096 s^{15}\!+\!2136820 s^{14} t\!+\!61039254 s^{13} t^2\\&
\!+\!1006276056 s^{12} t^3\!+\!9868826220 s^{11} t^4\!+\!63875343113 s^{10} t^5\!+\!292621754603 s^9 t^6\!+\!988998062953 s^8 t^7\\&
\!+\!2518818379737 s^7 t^8\!+\!4859851351942 s^6 t^9\!+\!7049500248014 s^5 t^{10}\!+\!7539507992112 s^4 t^{11}\\&
\!+\!5747250009888 s^3 t^{12}\!+\!2946149978112 s^2 t^{13}\!+\!908354248704 s t^{14}\!+\!127083479040 t^{15})\\&
\!-\!M_c^{10} (657624 s^{15}\!+\!39367890 s^{14} t\!+\!1104413029 s^{13} t^2\!+\!17344487058 s^{12} t^3\!+\!170832670892 s^{11} t^4\\&
\!+\!1146876637852 s^{10} t^5\!+\!5529671411978 s^9 t^6\!+\!19757278335172 s^8 t^7\!+\!53132548989137 s^7 t^8\\&
\!+\!107815796091060 s^6 t^9\!+\!163659426240484 s^5 t^{10}\!+\!182253522957504 s^4 t^{11}\!+\!143985271325760 s^3 t^{12}\\&
\!+\!76165483016448 s^2 t^{13}\!+\!24134155960320 s t^{14}\!+\!3456670629888 t^{15})\!+\!M_c^8 (33792 s^{16}\!+\!2774244 s^{15} t\\&
\!+\!97694544 s^{14} t^2\!+\!1994311862 s^{13} t^3\!+\!24356767073 s^{12} t^4\!+\!194789687794 s^{11} t^5\!+\!1093516581704 s^{10} t^6\\&
\!+\!4504581919900 s^9 t^7\!+\!13974266124379 s^8 t^8\!+\!33029076981316 s^7 t^9\!+\!59471161006872 s^6 t^{10}\\&
\!+\!80753245374608 s^5 t^{11}\!+\!81030643128336 s^4 t^{12}\!+\!58075814264832 s^3 t^{13}\!+\!28049364706176 s^2 t^{14}\\&
\!+\!8164110237696 s t^{15}\!+\!1080209571840 t^{16})
\bigr],
\addtocounter{equation}{1}
\tag{\theequation}
\end{align*}
}
{
\allowdisplaybreaks
\begin{align*}
F&^{(1)}_1[AVD_{cc}+AV\bar D_{\bar c\bar c}](s,t) = -\frac{262144}{27 M_c^5 s^7 \left(M_c^2-t\right)^5 \left(M_c^2-s-t\right)^5} \\ &
\!\times\!
\bigl[
1728 M_c^{30}\!-\!144 M_c^{28} (1393 s\!+\!144 t)\!+\!144 M_c^{26} (7517 s^2\!+\!16668 s t\!+\!792 t^2)\!-\!18 M_c^{24} (130927 s^3\\&
\!+\!658936 s^2 t\!+\!718992 s t^2\!+\!21120 t^3)\!+\!6 M_c^{22} (414903 s^4\!+\!3896270 s^3 t\!+\!9595884 s^2 t^2\!+\!6850464 s t^3\\&
\!+\!142560 t^4)\!-\!2 M_c^{20} (570966 s^5\!+\!10935641 s^4 t\!+\!49831179 s^3 t^2\!+\!80012160 s^2 t^3\!+\!41813496 s t^4\\&
\!+\!684288 t^5)\!+\!2 M_c^{18} (\!-\!70283 s^6\!+\!3939119 s^5 t\!+\!38212583 s^4 t^2\!+\!114578976 s^3 t^3\!+\!134518860 s^2 t^4\\&
\!+\!54136512 s t^5\!+\!798336 t^6)\!+\!M_c^{16} (514957 s^7\!+\!2926870 s^6 t\!-\!11814356 s^5 t^2\!-\!114600008 s^4 t^3\\&
\!-\!268330620 s^3 t^4\!-\!242157600 s^2 t^5\!-\!73652544 s t^6\!-\!1368576 t^7)\!-\!M_c^{14} (361295 s^8\!+\!4828723 s^7 t\\&
\!+\!24570531 s^6 t^2\!+\!51080312 s^5 t^3\!+\!31705580 s^4 t^4\!-\!8315448 s^3 t^5\!+\!8718192 s^2 t^6\!+\!17635968 s t^7\\&
\!-\!855360 t^8)\!+\!8 s t^3 (s\!+\!t)^3 (8 s^8\!+\!25 s^7 t\!+\!2536 s^6 t^2\!+\!21366 s^5 t^3\!+\!78759 s^4 t^4\!+\!157896 s^3 t^5\\&
\!+\!179640 s^2 t^6\!+\!108864 s t^7\!+\!27216 t^8)\!+\!M_c^{12} (151121 s^9\!+\!2725998 s^8 t\!+\!22467644 s^7 t^2\\&
\!+\!100637416 s^6 t^3\!+\!266920660 s^5 t^4\!+\!453877540 s^4 t^5\!+\!511993860 s^3 t^6\!+\!349531200 s^2 t^7\\&
\!+\!104894352 s t^8\!-\!380160 t^9)\!-\!M_c^2 s t^2 (s\!+\!t)^2 (320 s^9\!+\!3892 s^8 t\!+\!110625 s^7 t^2\!+\!1057188 s^6 t^3\\&
\!+\!4932499 s^5 t^4\!+\!13283958 s^4 t^5\!+\!21746386 s^3 t^6\!+\!21339360 s^2 t^7\!+\!11507112 s t^8\!+\!2612736 t^9)\\&
\!-\!M_c^{10} (42121 s^{10}\!+\!891469 s^9 t\!+\!9756298 s^8 t^2\!+\!61315158 s^7 t^3\!+\!234557673 s^6 t^4\!+\!572091236 s^5 t^5\\&
\!+\!907966580 s^4 t^6\!+\!913025712 s^3 t^7\!+\!525013056 s^2 t^8\!+\!129865536 s t^9\!-\!114048 t^{10})\!+\!M_c^8 (7984 s^{11}\\&
\!+\!172896 s^{10} t\!+\!2390911 s^9 t^2\!+\!20059044 s^8 t^3\!+\!102849261 s^7 t^4\!+\!335986166 s^6 t^5\!+\!718609852 s^5 t^6\\&
\!+\!1005256024 s^4 t^7\!+\!884461578 s^3 t^8\!+\!441939600 s^2 t^9\!+\!95071968 s t^{10}\!-\!20736 t^{11})\!+\!2 M_c^4 s t (360 s^{11}\\&
\!+\!7056 s^{10} t\!+\!133908 s^9 t^2\!+\!1462154 s^8 t^3\!+\!8991251 s^7 t^4\!+\!34117345 s^6 t^5\!+\!84461098 s^5 t^6\\&
\!+\!139249436 s^4 t^7\!+\!151574975 s^3 t^8\!+\!104472969 s^2 t^9\!+\!41237280 s t^{10}\!+\!7089912 t^{11})\!-\!M_c^6 (816 s^{12}\\&
\!+\!17956 s^{11} t\!+\!301806 s^{10} t^2\!+\!3524997 s^9 t^3\!+\!24696371 s^8 t^4\!+\!107402983 s^7 t^5\!+\!303244845 s^6 t^6\\&
\!+\!567265544 s^5 t^7\!+\!697527422 s^4 t^8\!+\!540613740 s^3 t^9\!+\!238658328 s^2 t^{10}\!+\!45603648 s t^{11}\!-\!1728 t^{12})
\addtocounter{equation}{1}
\tag{\theequation}
\bigr],
\end{align*}
\begin{align*}
F&^{(2)}_1[AVD_{cc}+AV\bar D_{\bar c\bar c}](s,t) = \frac{262144}{81 M_c^5 s^8 \left(M_c^2-t\right)^6 \left(M_c^2-s-t\right)^6} \\ &
\!\times\!
\bigl[
20221056 M_c^{36}\!-\!576 M_c^{34} (233941 s\!+\!561036 t)\!+\!144 M_c^{32} (2753877 s^2\!+\!14104544 s t\!+\!16667448 t^2)\\&
\!-\!4 M_c^{30} (170096945 s^3\!+\!1400883912 s^2 t\!+\!3527653392 s t^2\!+\!2738026368 t^3)\!+\!24 M_c^{28} (31650883 s^4\\&
\!+\!371666283 s^3 t\!+\!1502816442 s^2 t^2\!+\!2481655488 s t^3\!+\!1422434160 t^4)\!-\!M_c^{26} (570679427 s^5\\&
\!+\!9029032556 s^4 t\!+\!52252923804 s^3 t^2\!+\!138610384704 s^2 t^3\!+\!168906874176 s t^4\!+\!76170786048 t^5)\\&
\!+\!M_c^{24} (256191463 s^6\!+\!5795816012 s^5 t\!+\!46170706992 s^4 t^2\!+\!176966758928 s^3 t^3\!+\!347658174288 s^2 t^4\\&
\!+\!333497539584 s t^5\!+\!122937429888 t^6)\!-\!2 M_c^{22} (2466837 s^7\!+\!863671542 s^6 t\!+\!11311001039 s^5 t^2\\&
\!+\!62914230172 s^4 t^3\!+\!183012632274 s^3 t^4\!+\!286116482160 s^2 t^5\!+\!224984126880 s t^6\!+\!69253636608 t^7)\\&
\!+\!M_c^{20} (\!-\!91034264 s^8\!-\!813592713 s^7 t\!+\!48326175 s^6 t^2\!+\!28800780204 s^5 t^3\!+\!158775180088 s^4 t^4\\&
\!+\!405100852344 s^3 t^5\!+\!539810088048 s^2 t^6\!+\!356995482624 s t^7\!+\!90987152256 t^8)\!+\!M_c^{18} (75448045 s^9\\&
\!+\!1332147106 s^8 t\!+\!9770119165 s^7 t^2\!+\!38874308682 s^6 t^3\!+\!89128173899 s^5 t^4\!+\!110253341580 s^4 t^5\\&
\!+\!56052424932 s^3 t^6\!-\!1916424576 s^2 t^7\!+\!2804763456 s t^8\!+\!9582693120 t^9)\!-\!32 t^4 (s\!+\!t)^4 (104 s^{10}\\&
\!+\!349 s^9 t\!+\!38895 s^8 t^2\!+\!388900 s^7 t^3\!+\!1790450 s^6 t^4\!+\!4854528 s^5 t^5\!+\!8476608 s^4 t^6\!+\!9797760 s^3 t^7\\&
\!+\!7348320 s^2 t^8\!+\!3265920 s t^9\!+\!653184 t^{10})\!-\!M_c^{16} (35021597 s^{10}\!+\!833102650 s^9 t\!+\!8603539424 s^8 t^2\\&
\!+\!50889644964 s^7 t^3\!+\!193786066418 s^6 t^4\!+\!500022958888 s^5 t^5\!+\!890742644832 s^4 t^6\\&
\!+\!1097311247136 s^3 t^7\!+\!910938066192 s^2 t^8\!+\!463385184768 s t^9\!+\!108408972672 t^{10})\!+\!M_c^2 t^3 (s\!+\!t)^3\\&
\!\times\!(17984 s^{11}\!+\!225976 s^{10} t\!+\!8137786 s^9 t^2\!+\!91342653 s^8 t^3\!+\!513344601 s^7 t^4\!+\!1753334331 s^6 t^5\\&
\!+\!3950202209 s^5 t^6\!+\!6069862064 s^4 t^7\!+\!6365607660 s^3 t^8\!+\!4394297088 s^2 t^9\!+\!1806167808 s t^{10}\\&
\!+\!334430208 t^{11})\!+\!2 M_c^{14} (5369598 s^{11}\!+\!159532368 s^{10} t\!+\!2107768049 s^9 t^2\!+\!15931331624 s^8 t^3\\&
\!+\!77303868220 s^7 t^4\!+\!256070555832 s^6 t^5\!+\!594867075166 s^5 t^6\!+\!973395581112 s^4 t^7\\&
\!+\!1105034420970 s^3 t^8\!+\!832224987792 s^2 t^9\!+\!374238168480 s t^{10}\!+\!75776553216 t^{11})\!-\!M_c^4 t^2 (s\!+\!t)^2\\&
\!\times\!(43888 s^{12}\!+\!903104 s^{11} t\!+\!22877182 s^{10} t^2\!+\!282654857 s^9 t^3\!+\!1903612637 s^8 t^4\!+\!8005381771 s^7 t^5\\&
\!+\!22622947685 s^6 t^6\!+\!44524612444 s^5 t^7\!+\!61674802264 s^4 t^8\!+\!59340363672 s^3 t^9\!+\!37899841968 s^2 t^{10}\\&
\!+\!14458383360 s t^{11}\!+\!2488195584 t^{12})\!-\!M_c^{12} (2248090 s^{12}\!+\!79574449 s^{11} t\!+\!1317350515 s^{10} t^2\\&
\!+\!12475000442 s^9 t^3\!+\!74917211664 s^8 t^4\!+\!304907732366 s^7 t^5\!+\!873255100426 s^6 t^6\\&
\!+\!1787109792136 s^5 t^7\!+\!2607119746632 s^4 t^8\!+\!2654336950408 s^3 t^9\!+\!1794055635504 s^2 t^{10}\\&
\!+\!722668267008 s t^{11}\!+\!130908645504 t^{12})\!+\!M_c^{10} (308488 s^{13}\!+\!12561878 s^{12} t\!+\!259953855 s^{11} t^2\\&
\!+\!3131090454 s^{10} t^3\!+\!23484836972 s^9 t^4\!+\!117277669268 s^8 t^5\!+\!408772826718 s^7 t^6\\&
\!+\!1021374753100 s^6 t^7\!+\!1847123871971 s^5 t^8\!+\!2403577776876 s^4 t^9\!+\!2196880552908 s^3 t^{10}\\&
\!+\!1338590478528 s^2 t^{11}\!+\!487521423936 s t^{12}\!+\!80079439104 t^{13})\!+\!2 M_c^6 t (27696 s^{14}\!+\!788592 s^{13} t\\&
\!+\!18122252 s^{12} t^2\!+\!256127153 s^{11} t^3\!+\!2159709194 s^{10} t^4\!+\!11773460135 s^9 t^5\!+\!44215597904 s^8 t^6\\&
\!+\!118957864637 s^7 t^7\!+\!234050783174 s^6 t^8\!+\!338739874305 s^5 t^9\!+\!357552284964 s^4 t^{10}\\&
\!+\!268091824970 s^3 t^{11}\!+\!135283963248 s^2 t^{12}\!+\!41152829472 s t^{13}\!+\!5692280832 t^{14})\!-\!M_c^8 (21360 s^{14}\\&
\!+\!1150240 s^{13} t\!+\!29400040 s^{12} t^2\!+\!473008826 s^{11} t^3\!+\!4622824387 s^{10} t^4\!+\!29030729246 s^9 t^5\\&
\!+\!124308726592 s^8 t^6\!+\!377796208660 s^7 t^7\!+\!833064465237 s^6 t^8\!+\!1342006821284 s^5 t^9\\&
\!+\!1567237665232 s^4 t^{10}\!+\!1293304522224 s^3 t^{11}\!+\!714916476432 s^2 t^{12}\!+\!237204615168 s t^{13}\\&
\!+\!35637528960 t^{14})
\addtocounter{equation}{1}
\tag{\theequation}
\bigr],
\end{align*}
\begin{align*}
F&^{(3)}_1[AVD_{cc}+AV\bar D_{\bar c\bar c}](s,t) = \frac{262144}{243 M_c^5 s^{10} \left(M_c^2-t\right)^7 \left(M_c^2-s-t\right)^7} \\
&
\!\times\!\bigl[1741824000 M_c^{44}\!-\!13824 M_c^{42} (1024717 s\!+\!2520000 t)\!+\!3456 M_c^{40} (15157181 s^2\!+\!78409048 s t\\&
\!+\!95243904 t^2)\!-\!768 M_c^{38} (152602537 s^3\!+\!1244415438 s^2 t\!+\!3174024582 s t^2\!+\!2539071360 t^3)\\&
\!+\!48 M_c^{36}(3709614441 s^4\!+\!42282452456 s^3 t\!+\!169803625392 s^2 t^2\!+\!284772736512 s t^3\!+\!168755166720 t^4)\\&
\!-\!8 M_c^{34} (24326003617 s^5\!+\!362802284320 s^4 t\!+\!2037040979064 s^3 t^2\!+\!5374375596864 s^2 t^3\\&
\!+\!6667854384384 s t^4\!+\!3121487953920 t^5)\!+\!16 M_c^{32} (9762324075 s^6\!+\!183890145551 s^5 t\\&
\!+\!1354940349368 s^4 t^2\!+\!5002564231056 s^3 t^3\!+\!9760290607872 s^2 t^4\!+\!9555769117440 s t^5\\&
\!+\!3679463854080 t^6)\!-\!4 M_c^{30} (22457637797 s^7\!+\!532549281496 s^6 t\!+\!4996889068172 s^5 t^2\\&
\!+\!24325079883568 s^4 t^3\!+\!66552894119904 s^3 t^4\!+\!102498878252160 s^2 t^5\!+\!82476996844032 s t^6\\&
\!+\!26843207860224 t^7)\!+\!M_c^{28} (31939769109 s^8\!+\!1031865139936 s^7 t\!+\!12519019456864 s^6 t^2\\&
\!+\!78810576693872 s^5 t^3\!+\!286787234374848 s^4 t^4\!+\!622093712687616 s^3 t^5\!+\!788656117759488 s^2 t^6\\&
\!+\!536192491511808 s t^7\!+\!150284156682240 t^8)\!-\!M_c^{26} (318053197 s^9\!+\!218468720272 s^8 t\\&
\!+\!4411307171664 s^7 t^2\!+\!38715857406848 s^6 t^3\!+\!189114078871984 s^5 t^4\!+\!556628554281920 s^4 t^5\\&
\!+\!1004225800991616 s^3 t^6\!+\!1080262099763712 s^2 t^7\!+\!632454318425088 s t^8\!+\!154289933844480 t^9)\\&
\!+\!M_c^{24} (\!-\!8778960932 s^{10}\!-\!134452810824 s^9 t\!-\!488149274250 s^8 t^2\!+\!4001218034320 s^7 t^3\\&
\!+\!48976432024272 s^6 t^4\!+\!230412433439024 s^5 t^5\!+\!608946149725312 s^4 t^6\!+\!966382124532480 s^3 t^7\\&
\!+\!907484756435712 s^2 t^8\!+\!461490695018496 s t^9\!+\!97128844001280 t^{10})\!+\!M_c^{22} (7078070576 s^{11}\\&
\!+\!173778399959 s^{10} t\!+\!1874571584019 s^9 t^2\!+\!11483453562160 s^8 t^3\!+\!43164708110292 s^7 t^4\\&
\!+\!100643321821024 s^6 t^5\!+\!138667494855120 s^5 t^6\!+\!93684817569344 s^4 t^7\!+\!3332723177088 s^3 t^8\\&
\!-\!24935540497920 s^2 t^9\!+\!968161379328 s t^{10}\!+\!7021362216960 t^{11})\!-\!16 t^5 (s\!+\!t)^5 (3960 s^{12}\!+\!14151 s^{11} t\\&
\!+\!1702238 s^{10} t^2\!+\!20378278 s^9 t^3\!+\!120761039 s^8 t^4\!+\!456128424 s^7 t^5\!+\!1201156440 s^6 t^6\!+\!2269682496 s^5 t^7\\&
\!+\!3068773200 s^4 t^8\!+\!2893812480 s^3 t^9\!+\!1805006592 s^2 t^{10}\!+\!668860416 s t^{11}\!+\!111476736 t^{12})\!-\!M_c^{20}\\&
\!\times\! (3405990647 s^{12}\!+\!104523726104 s^{11} t\!+\!1437801582469 s^{10} t^2\!+\!11759612667954 s^9 t^3\\&
\!+\!63654208362929 s^8 t^4\!+\!240386484084080 s^7 t^5\!+\!652169234475584 s^6 t^6\!+\!1289432039042896 s^5 t^7\\&
\!+\!1857555696533856 s^4 t^8\!+\!1912110070725888 s^3 t^9\!+\!1336908589885440 s^2 t^{10}\!+\!568038597894144 s t^{11}\\&
\!+\!110118364102656 t^{12})\!+\!4 M_c^2 t^4 (s\!+\!t)^4 (91504 s^{13}\!+\!1159496 s^{12} t\!+\!51949345 s^{11} t^2\!+\!687364992 s^{10} t^3\\&
\!+\!4769853365 s^9 t^4\!+\!21399318662 s^8 t^5\!+\!67686740542 s^7 t^6\!+\!156321971272 s^6 t^7\!+\!265402870978 s^5 t^8\\&
\!+\!327237185520 s^4 t^9\!+\!284374491696 s^3 t^{10}\!+\!164763942912 s^2 t^{11}\!+\!57042994176 s t^{12}\!+\!8918138880 t^{13})\\&
\!+\!M_c^{18} (1131091631 s^{13}\!+\!42014573246 s^{12} t\!+\!693307252728 s^{11} t^2\!+\!6787879263196 s^{10} t^3\\&
\!+\!44334703856018 s^9 t^4\!+\!204596890047160 s^8 t^5\!+\!687431049901512 s^7 t^6\!+\!1704996711408576 s^6 t^7\\&
\!+\!3124052755731456 s^5 t^8\!+\!4176368330021184 s^4 t^9\!+\!3956371217584896 s^3 t^{10}\!+\!2511745562866176 s^2 t^{11}\\&
\!+\!956202781427712 s t^{12}\!+\!164551924776960 t^{13})\!-\!M_c^4 t^3 (s\!+\!t)^3 (962656 s^{14}\!+\!21094520 s^{13} t\\&
\!+\!691108948 s^{12} t^2\!+\!10002353127 s^{11} t^3\!+\!80665258449 s^{10} t^4\!+\!426581637437 s^9 t^5\!+\!1603433166215 s^8 t^6\\&
\!+\!4449490702576 s^7 t^7\!+\!9237022428688 s^6 t^8\!+\!14298644494848 s^5 t^9\!+\!16229093115728 s^4 t^{10}\\&
\!+\!13078157302656 s^3 t^{11}\!+\!7067907479808 s^2 t^{12}\!+\!2293062524928 s t^{13}\!+\!337147453440 t^{14})\!-\!2 M_c^{16}\\&
\!\times\!(133029237 s^{14}\!+\!5947101882 s^{13} t\!+\!117421358999 s^{12} t^2\!+\!1355188930116 s^{11} t^3\!+\!10363630842574 s^{10} t^4\\&
\!+\!56081817619836 s^9 t^5\!+\!222407868023894 s^8 t^6\!+\!657773336913072 s^7 t^7\!+\!1458291508962840 s^6 t^8\\&
\!+\!2410239762599576 s^5 t^9\!+\!2920175104142912 s^4 t^{10}\!+\!2512162544509056 s^3 t^{11}\\&
\!+\!1449370729677312 s^2 t^{12}\!+\!501693012043776 s t^{13}\!+\!78585246351360 t^{14})\!+\!M_c^6 t^2 (s\!+\!t)^2 (1391744 s^{15}\\&
\!+\!43117128 s^{14} t\!+\!1279151434 s^{13} t^2\!+\!20465640103 s^{12} t^3\!+\!192094141903 s^{11} t^4\!+\!1196102161809 s^{10} t^5\\&
\!+\!5318240639251 s^9 t^6\!+\!17564547014872 s^8 t^7\!+\!43863878016444 s^7 t^8\!+\!83088452331520 s^6 t^9\\&
\!+\!118407481083504 s^5 t^{10}\!+\!124541123134400 s^4 t^{11}\!+\!93507020744832 s^3 t^{12}\!+\!47299950093312 s^2 t^{13}\\&
\!+\!14419185974784 s t^{14}\!+\!1998499184640 t^{15})\!+\!M_c^{14} (43621482 s^{15}\!+\!2343138771 s^{14} t\\&
\!+\!56199599441 s^{13} t^2\!+\!773612810138 s^{12} t^3\!+\!6932647567340 s^{11} t^4\!+\!43593912548178 s^{10} t^5\\&
\!+\!200838117586542 s^9 t^6\!+\!693892463426240 s^8 t^7\!+\!1816416867876372 s^7 t^8\!+\!3603519546818400 s^6 t^9\\&
\!+\!5370130545307120 s^5 t^{10}\!+\!5898572062338496 s^4 t^{11}\!+\!4619743847396736 s^3 t^{12}\\&
\!+\!2435030828186112 s^2 t^{13}\!+\!772528110925824 s t^{14}\!+\!111261585899520 t^{15})\!-\!M_c^{12} (4569808 s^{16}\\&
\!+\!303380716 s^{15} t\!+\!9094156671 s^{14} t^2\!+\!154232294898 s^{13} t^3\!+\!1654322273320 s^{12} t^4\\&
\!+\!12195176483488 s^{11} t^5\!+\!65211887760950 s^{10} t^6\!+\!261153355185708 s^9 t^7\!+\!796490061350461 s^8 t^8\\&
\!+\!1861164352695808 s^7 t^9\!+\!3323329278341024 s^6 t^{10}\!+\!4486352440347568 s^5 t^{11}\\&
\!+\!4487438605163072 s^4 t^{12}\!+\!3215047033035264 s^3 t^{13}\!+\!1556602532015616 s^2 t^{14}\\&
\!+\!455363166265344 s t^{15}\!+\!60694275317760 t^{16})\!-\!2 M_c^8 t (473008 s^{17}\!+\!22997364 s^{16} t\!+\!723877564 s^{15} t^2\\&
\!+\!13517149022 s^{14} t^3\!+\!154742601512 s^{13} t^4\!+\!1189403439533 s^{12} t^5\!+\!6565310531804 s^{11} t^6\\&
\!+\!27133825223148 s^{10} t^7\!+\!85972286205904 s^9 t^8\!+\!211182065007357 s^8 t^9\!+\!403046816578584 s^7 t^{10}\\&
\!+\!594612672095720 s^6 t^{11}\!+\!669840935465464 s^5 t^{12}\!+\!564179282090816 s^4 t^{13}\!+\!343245181246848 s^3 t^{14}\\&
\!+\!142249221368640 s^2 t^{15}\!+\!35889377398272 s t^{16}\!+\!4155365007360 t^{17})\!+\!M_c^{10} (231552 s^{17}\\&
\!+\!23087576 s^{16} t\!+\!898504584 s^{15} t^2\!+\!19979428298 s^{14} t^3\!+\!269237598605 s^{13} t^4\!+\!2395515473146 s^{12} t^5\\&
\!+\!15079912639380 s^{11} t^6\!+\!70246320893068 s^{10} t^7\!+\!248576830505007 s^9 t^8\!+\!677016721065728 s^8 t^9\\&
\!+\!1424186961493664 s^7 t^{10}\!+\!2304170990875904 s^6 t^{11}\!+\!2833619756443696 s^5 t^{12}\\&
\!+\!2594347606941888 s^4 t^{13}\!+\!1708831468731264 s^3 t^{14}\!+\!763746183277056 s^2 t^{15}\\&
\!+\!207045959569920 s t^{16}\!+\!25667685679104 t^{17})
\addtocounter{equation}{1}
\tag{\theequation}
\bigr].
\end{align*}
}

\end{document}